\documentclass[prb,aps,amsmath,amssymb,amsfonts,superscriptaddress,showpacs,floatfix,twocolumn]{revtex4}
\usepackage[dvips,usenames]{color}
\usepackage{amsmath}
\usepackage{float}
\usepackage{amssymb}
\usepackage{epsfig}
\usepackage{epsf}
\usepackage{array}
\usepackage{graphicx}
\usepackage{epstopdf}
\usepackage{subfigure}
\usepackage{bbm}
\usepackage{hyperref}

\usepackage{color}
\def\Im{\hbox{Im}}

\begin{document}

\title{Momentum-space anisotropy and pseudogaps: a comparative cluster dynamical mean-field analysis of the doping-driven metal-insulator transition in the two-dimensional Hubbard model}

\author{E. Gull}
\affiliation{Department of Physics, Columbia University, 538 West 120th Street, New York, New York 10027, USA}

\author{M. Ferrero}
\affiliation{Centre de Physique Th\'eorique, \'Ecole Polytechnique, CNRS, 91128 Palaiseau Cedex, France}

\author{O. Parcollet}
\affiliation{Institut de Physique Th\'eorique, CEA/DSM/IPhT-CNRS/URA 2306, CEA-Saclay, F-91191 Gif-sur-Yvette, France}

\author{A. Georges}
\affiliation{Centre de Physique Th\'eorique, \'Ecole Polytechnique, CNRS, 91128 Palaiseau Cedex, France}
\affiliation{Coll\`ege de France, 11 Place Marcelin Berthelot, 75231 Paris Cedex 05, France}

\author{A. J. Millis}
\affiliation{Department of Physics, Columbia University, 538 West 120th Street, New York, New York 10027, USA}

\begin{abstract}
Cluster dynamical mean-field calculations based on  $2$-, $4$-, $8$- and $16$-site clusters  are used to analyze  the doping-driven metal-insulator transition in the two-dimensional Hubbard model. Comparison of results obtained on different clusters enables a determination of  those aspects of the physics that are common to all clusters and permits identification of artifacts associated with particular cluster geometries. A modest particle-hole asymmetry in the underlying band structure is shown to lead to qualitatively different behavior on the hole-doped side than on the electron-doped side.   For particle-hole asymmetry of the sign and magnitude appropriate to high-$T_c$ cuprates, the approach to the insulator from the hole-doping side is found to  proceed in two stages from a high-doping region where the  properties are those of a Fermi liquid with moderately renormalized parameters and very weak momentum dependence. As doping is reduced the system first enters an intermediate doping regime where the Fermi-liquid renormalizations are larger and the electron self-energy varies significantly around the Fermi surface and then passes  to  a small doping regime characterized by a gap on some parts of the Fermi surface but gapless behavior in other parts. On the electron-doped side the partially gapped regime does not occur, and  the momentum dependence of the electron self-energy is less pronounced.  Implications for the high-$T_c$ cuprates and for  the use of cluster dynamical mean-field methods in wider classes of problems are discussed.
\end{abstract}

\date{\today}

\pacs{71.10.Fd, 71.27.+a,71.30.+h,74.72.-h}

\maketitle
\section{Introduction}
\label{sec:intro}

The evolution of a Mott (correlation-driven) insulating state as charge carriers are induced by doping is one of the basic questions in condensed-matter physics. It is  relevant to the behavior of wide classes of materials \cite{Imada98} and is of particular importance in the context of high-temperature copper-oxide superconductivity, where following an early insight of Anderson \cite{Anderson87} the materials are generally accepted to be doped Mott insulators, with the combination of  ``Mott'' correlations and (quasi) two-dimensionality giving rise to the novel properties.  

A crucial insight into the physics of doped Mott insulators was provided by  Brinkman and Rice \cite{Brinkman70} who argued that the low-temperature properties of the doped Mott insulator were in effect those of a Fermi liquid with a quasiparticle mass which diverged  as the Mott insulating state was approached. The Brinkman-Rice picture provides a reasonable description of data in many materials \cite{Imada98} and has been refined theoretically over many years. In particular the development of the single-site dynamical mean-field theory \cite{Georges96} provided a precise theoretical framework within which these results can be derived. The key physical  assumptions required for Brinkman-Rice behavior are now understood to be the presence of strong correlations and the  locality of correlation effects. The mathematical expression of the latter assumption is the momentum independence of the electron self-energy. The locality  assumption becomes strictly valid for classes of  lattice models  in a limit of infinite  lattice coordination number\cite{Metzner89} and appears to provide a reasonable approximation to the behavior of many three-dimensional materials.\cite{Imada98} However, the momentum-independent self-energy approximation is likely to be less accurate for materials, such as high-$T_c$ superconductors, where the electronic properties are two-dimensional. 

Indeed, it was recognized  soon after the discovery of high-$T_c$ superconductivity in the CuO$_2$ perovskites that the  materials required a description which went beyond the Brinkman-Rice/single-site dynamical mean-field theory approach. The evidence has become stronger over the years and it is beyond the scope of this paper to review the large experimental literature demonstrating this point. We do recall here the results of angle-resolved photoemission and related transport  experiments which clearly illustrate this point.  Studies of  hole-doped cuprate materials with very high dopings (above the doping which maximizes the superconducting transition temperature)  indicate a quasiparticle lifetime and velocity
renormalization  which are  nearly isotropic around the Fermi surface,\cite{Yusof02} consistent with inferences from magnetotransport.\cite{Abdel-Jawad07,French09}  As the doping is reduced toward the insulating phase the behavior changes. Photoemission measurements on materials with ``optimal'' doping levels (near the carrier concentration which maximizes the superconducting transition temperature) reveal an electron lifetime which varies dramatically around the Fermi surface,\cite{Valla99} again consistent with inferences drawn from interpretations of the interplane conductivity \cite{Ioffe98} and from in-plane magnetoresistance measurements.\cite{Abdel-Jawad07} Measurements on ``underdoped'' cuprates (those with carrier concentrations even  closer to the insulating phase) reveal a ``pseudogap'' (reduction in electronic density of states) for momenta near the $(0,\pi)/(\pi,0)$ points of the Brillouin zone but no pseudogap for states near the zone-diagonal (near $(\pi/2,\pi/2)$).\cite{Loeser96,Ding96,Shen03} Thus optimally doped and underdoped cuprate materials exhibit a self-energy with a strong momentum dependence, inconsistent with the Brinkman-Rice/single-site dynamical mean-field approach. 

On the theoretical side the importance of going beyond the Brinkman-Rice description was  also recognized early on. For example, the various forms of resonating valence bond (``RVB'')  theories take intersite correlations explicitly into  account by expansion around a specific mean-field approximation and  were shown to lead to a non-Brinkman-Rice doping dependence of the electron effective mass and other Fermi liquid parameters \cite{Grilli90a,Grilli90b} and to a pseudogap \cite{Wen96} (for a recent review see Ref.~\onlinecite{Lee07}). Other authors  have addressed the issue in the context of analytical calculations based on the assumed importance of antiferromagnetic \cite{Kampf90,Vilk97,Abanov01,Kyung04} or charge-density wave \cite{Castellani95} correlations. These, and other semi-analytic calculations, while demonstrating the importance of antiferromagnetic correlations for the electron-doped cuprates, have not led to a consensus regarding the physics of the hole-doped cuprates, in part because they are based on approximations  which are uncontrolled (or are controlled in limits which are not clearly relevant to the actual materials) and more importantly because they are based on assumptions  about which correlations are physically relevant and which may be neglected.  The appropriateness of the  underlying assumptions about which physics to include have been the subject of debate.

Over the last decade the development of ``cluster'' dynamical mean-field methods \cite{Hettler98,Hettler00,Lichtenstein00,Kotliar01} has opened up a very promising new line of attack on the problem. These methods obtain an approximate solution of the full many-body problem in terms of the solution of an auxiliary $N$-site quantum impurity model coupled with a self-consistency condition.  For a review, see Ref.~\onlinecite{Maier05} and Refs.~\onlinecite{Kotliar06,Tremblay06}. $N=1$ corresponds to single-site dynamical mean-field theory with a momentum-independent self-energy; clusters of size $N>1$   allow for some momentum dependence of the self-energy and thus   enable the study  of deviations from Brinkman-Rice behavior. As $N\rightarrow\infty$ one recovers the full model; however, the computational expense rises rapidly as the interaction strength and cluster size increase.  An advantage of the methods is that no explicit assumption is made about the important of one kind of electronic correlation (spin density, charge density, RVB) relative to another, but the possibility of potential biases associated with choice of cluster is an important issue which this paper aims to address. 

\begin{figure}[t]
\includegraphics[width=0.85\columnwidth]{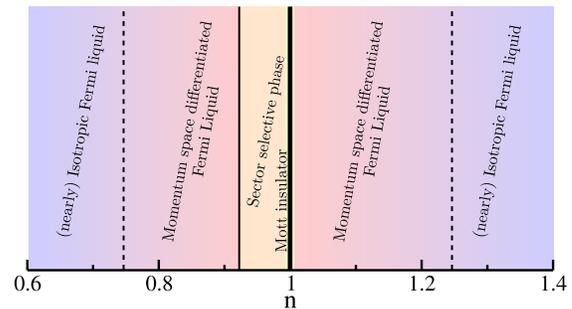}
\caption{\label{fig:Cartoon}Qualitative sketch of doping regimes for parameters considered in this paper.
}

\end{figure}
\begin{figure*}[t]
\includegraphics[width=0.19\textwidth]{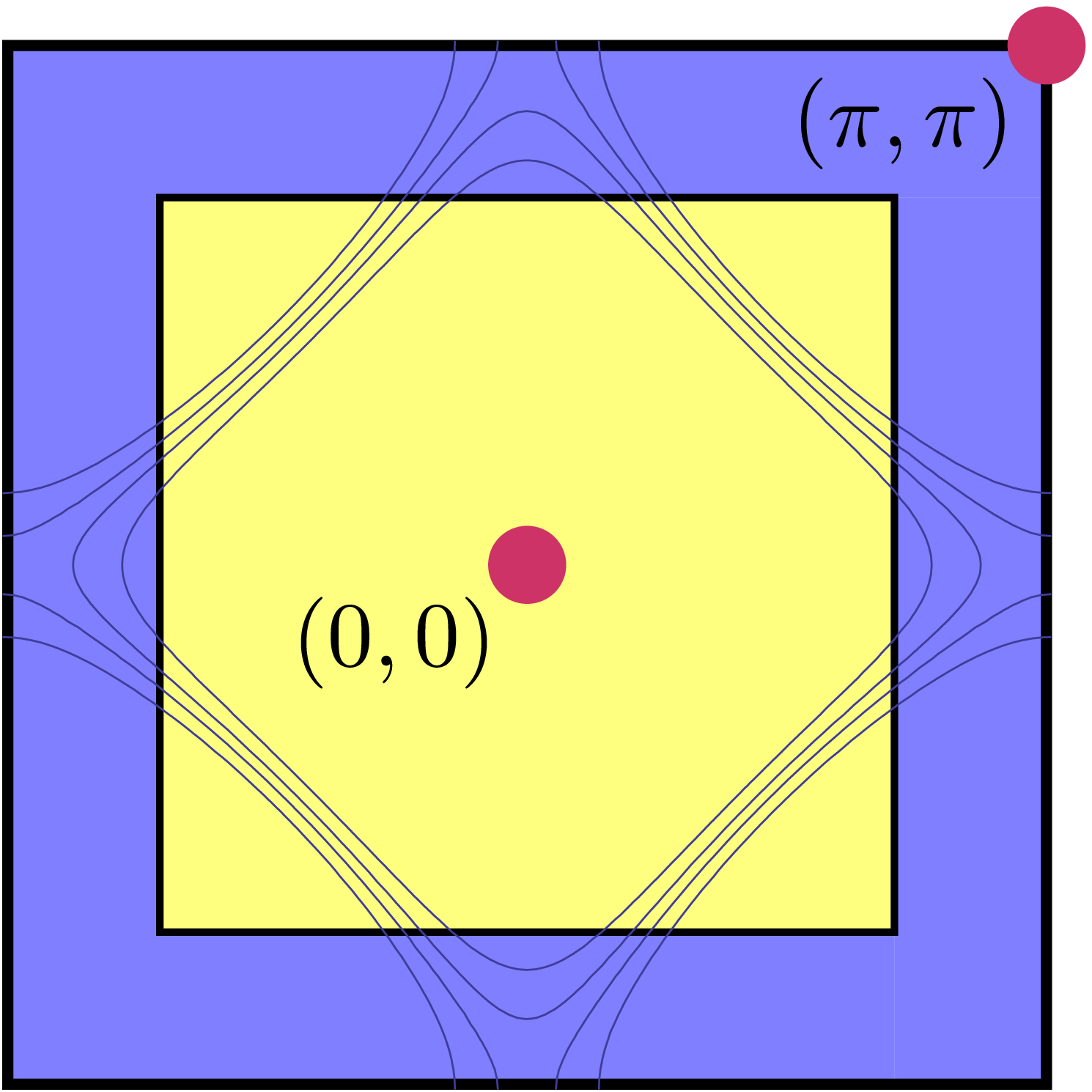}
\includegraphics[width=0.19\textwidth]{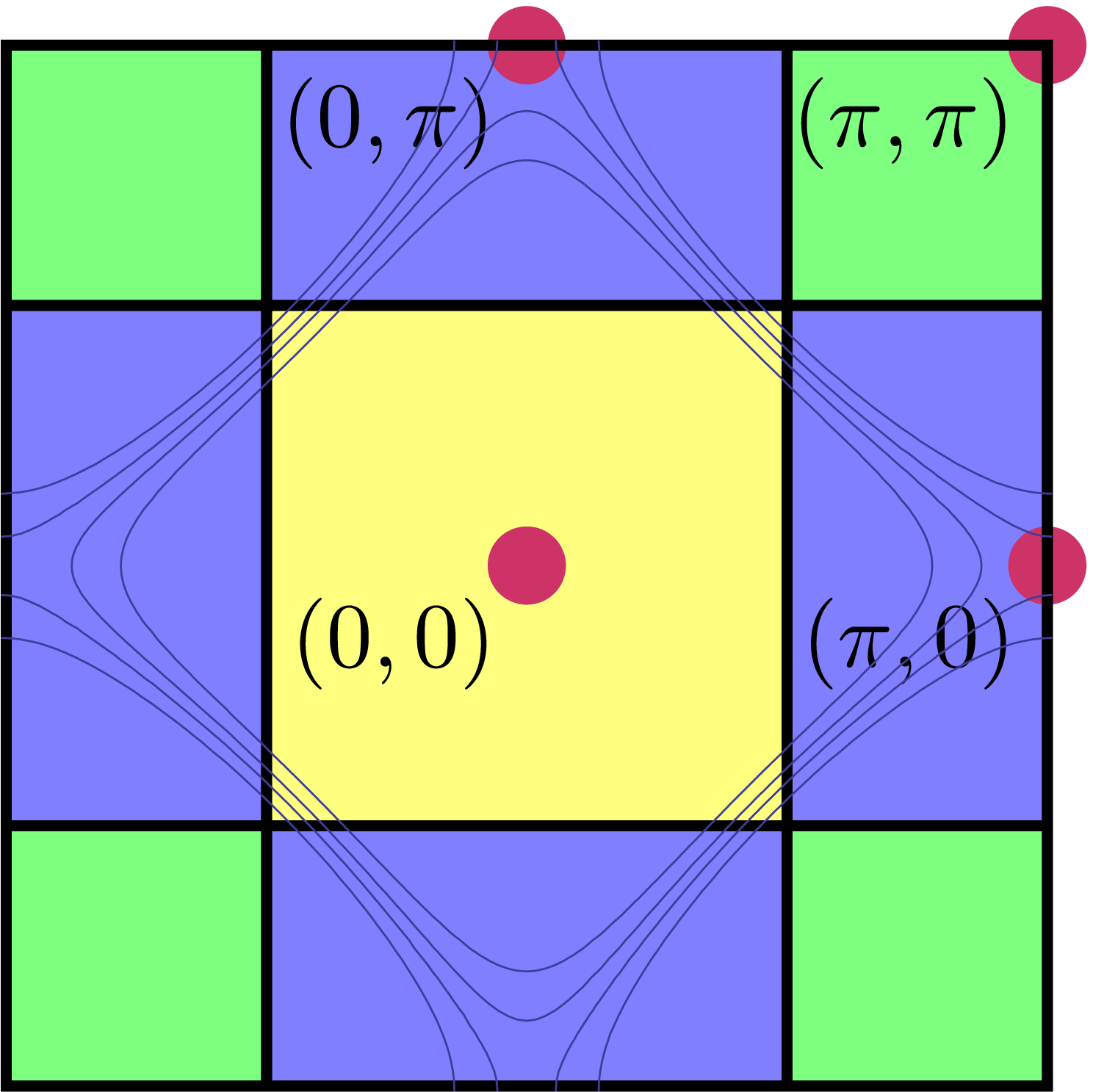}
\includegraphics[width=0.19\textwidth]{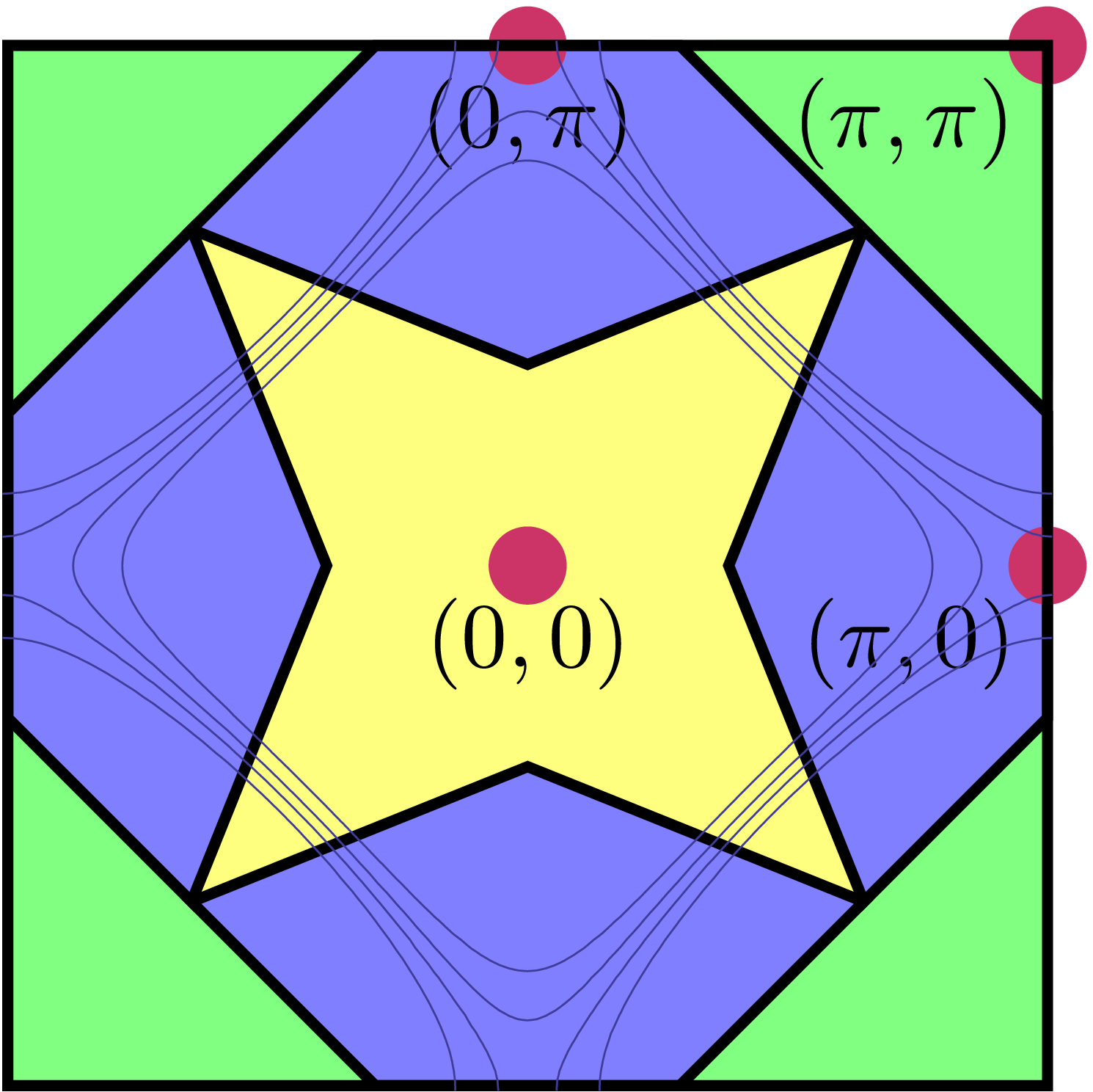}
\includegraphics[width=0.19\textwidth]{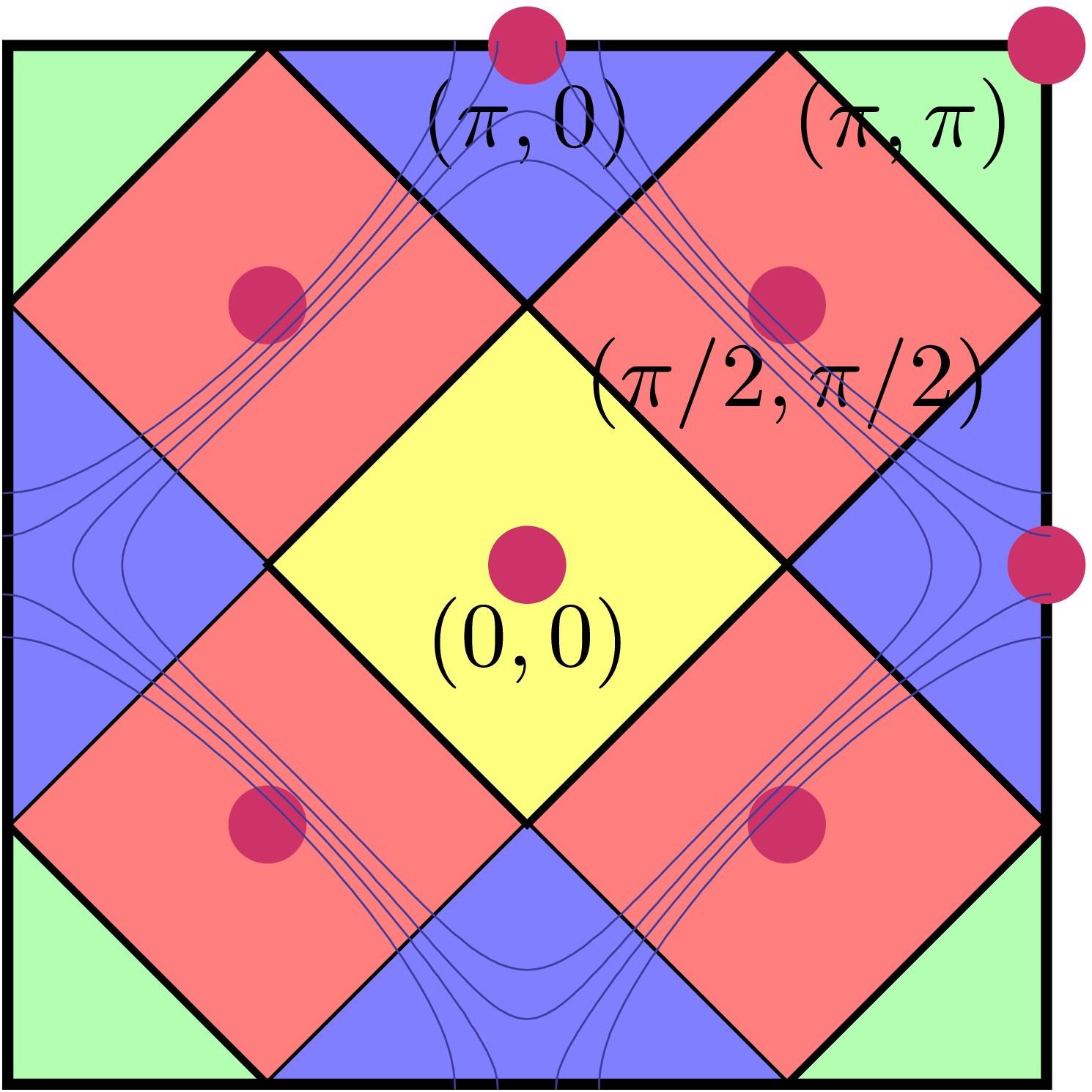}
\includegraphics[width=0.19\textwidth]{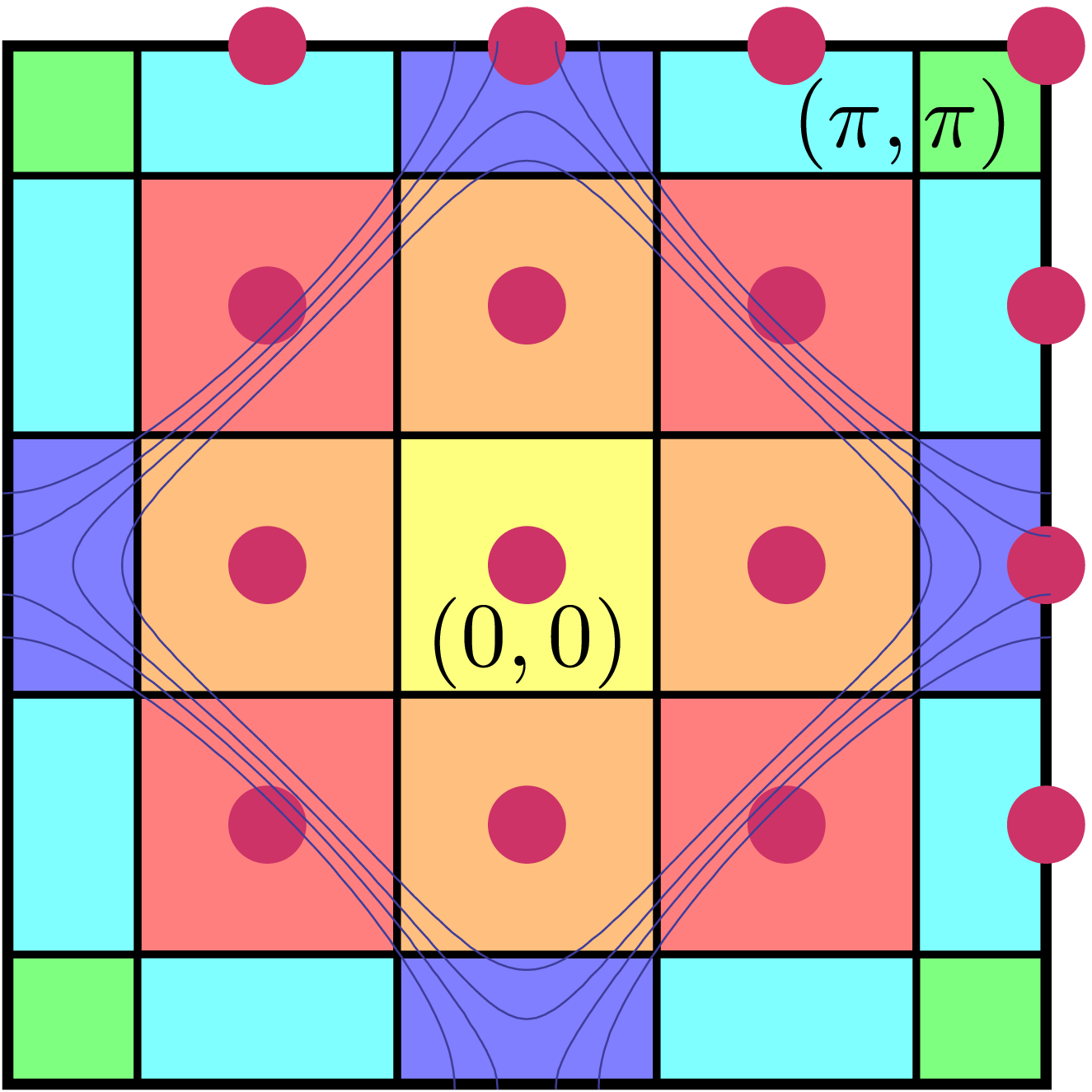}\\
\caption{(Color online) Momentum-space tiling used to define cluster approximations studied here: $2$-site (leftmost panel), $4$-site with standard patching (second from left), $4$-site with alternative patching ($4^*$), (central panel), $8$-site (second from right), and $16$-site (rightmost panel).  Momentum-space patches indicated by shaded regions; electron self-energy is independent of momentum within a patch but may vary from patch to patch. Dots (red) represent the $K$ points in reciprocal space associated to the patches in the DCA construction (see text). Thin lines : Fermi surfaces for the noninteracting system with $t'=-0.15t$ for half filling and hole dopings of 10\%, 20\%, and 30\%. All clusters have an inner patch around $(0,0)$ (yellow) and an outer patch around $(\pi,\pi)$ (green). Clusters with four or more sites also have an antinodal patch at $(\pi,0)$ and symmetry-related points (blue), clusters with eight or more sites have a nodal patch ($(\pi/2,\pi/2)$, red). The 16-site cluster has two additional independent momentum sectors, around $(\pi/2,0)$ (orange) and around $(3 \pi/2, \pi/2)$ (cyan). All clusters have the full point-group symmetry of the lattice. 
}
\label{fig:clusters}
\end{figure*}

Important results obtained by cluster dynamical mean-field methods  have included the demonstration that in an appropriate doping and interaction range the two-dimensional Hubbard model can exhibit a pseudogap,\cite{Huscroft01,Macridin06}
``Fermi arcs'',\cite{Parcollet04,Civelli05,Berthod06,Kyung06,Stanescu06} and a variation in electronic properties around the Fermi surface (``nodal-antinodal differentiation'').\cite{Parcollet04,Civelli05,Macridin06,Ferrero09,Ferrero09B,Ferrero10} The methods have been shown to yield a multistage approach to the Mott transition with the insulating phase being separated from the weakly correlated metallic phase by a ``sector-selective'' phase where some regions of the Brillouin zone are gapped and others are not \cite{Werner09,Gull09,Ferrero09,Ferrero09B} and a number of physical properties were shown to be in good agreement with experiment.\cite{Lin10,Ferrero10,Ferrero09B}

Despite these successes, uncertainties remain. Most cluster dynamical mean-field papers analyze one specific cluster.  Comparison of results obtained on clusters of different sizes has been undertaken only in a few special cases,\cite{Jarrell01,Moukouri01,Maier05_dwave,Kent05,Kozik10} mostly not directly relevant to the question of the doping-driven Mott transition.  It has therefore not been clear which results are due to specific properties of clusters and which results are representative of the physics of the full model.  More generally, cluster dynamical mean-field calculations test the limits of present day computational abilities so that compromises are required between cluster size and the ranges of  temperature, interaction strength and carrier concentration to be studied. Little information is available in the literature to guide the choices which must be made.

This paper has two main goals: to clarify the physics of the doping-driven Mott transition in two dimensions  by identifying  the robust physical features  which can now be considered as established from cluster dynamical mean-field theory and to   identify the differences between different cluster sizes and geometries.   We present a global examination of the one-electron properties of the doped Mott insulator, using clusters of all feasible sizes from $2$ up to $16$ sites, respecting the lattice symmetry. The investigation  is made possible by progress in algorithms, which have made the computations much more efficient, enabling surveys of wide ranges of parameter space for many clusters.\cite{Gull08_ctaux,Alvarez08} We describe the electronic properties in detail and show that the phase diagram and physical properties are to a surprising extent robust against choice of cluster size and geometry. Where differences occur, the features of the cluster which cause them are determined. Our results define the  current frontier of the field, given present  computational capabilities, and call for a new generation of theoretical developments aiming at improving momentum-space resolution.

 While the various aspects of the doping-dependent phase diagram of the two-dimensional Hubbard model have been noted in various ways  in the cluster dynamical mean-field literature, the generality of the results and their robustness to choice of cluster have not been previously appreciated.  The comparison of results for different sized clusters  clearly demonstrates that  the essentials of the carrier concentration dependence of physical properties of a doped Mott insulator are as sketched in Fig.~\ref{fig:Cartoon}. Far from the insulating state, the properties are those of a moderately correlated Fermi liquid.  Moreover, the momentum dependence of the renormalizations is very weak: the properties are described well by single-site dynamical mean-field theory, as previously noted, e.g., in Refs.~\onlinecite{Parcollet04} and \onlinecite{Civelli05}. We refer to this regime as the {\em isotropic Fermi liquid}. [Note that ``isotropic'' here means isotropic scattering properties (self-energy) along the Fermi surface, but the Fermi surface is not circular.] As the doping is decreased towards the $n=1$ insulating state the system enters an intermediate doping regime where the low-temperature behavior is still described by Fermi-liquid theory, but the Fermi liquid is characterized by a strong momentum dependence of the self-energies with the self-energies being largest near the zone corner $(0,\pi)/(\pi,0)$ points and smallest near the zone diagonal $(\pm \pi/2,\pm \pi/2)$ regions of momentum-space.  We refer to this as the regime of {\em momentum-space differentiation}. The change between the isotropic and momentum-space differentiated Fermi-liquid regimes is not characterized by any order parameter and we believe it to be a crossover, not a transition, but the doping at which the change occurs is surprisingly sharply defined and is indicated by dashed lines in Fig.~\ref{fig:Cartoon}.

As the doping is decreased yet further, a non-Fermi-liquid regime appears  on the hole-doping side  but not on the electron-doping side (for the moderate particle-hole asymmetry considered here). In the non-Fermi-liquid regime, regions of momentum-space near $(0,\pi)/(\pi,0)$ acquire an interaction-induced gap, while the zone diagonal regions of momentum-space remain gapless. We refer to this regime as the {\em sector-selective} regime. The boundary between the regime of momentum-space differentiation and the sector-selective regime is indicated by a light solid line in Fig.~\ref{fig:Cartoon}. Finally at doping $n=1$ the system is in the Mott insulating phase. 

The remainder of the paper is organized as follows. In Sec.~\ref{sec:overview} we summarize the general features of the doping-driven Mott transition, define the model to be studied and the questions to be considered and outline the theoretical approach. In Sec.~\ref{sec:regimes}  we demonstrate the existence of different doping regimes and how they  appear in
the different cluster calculations. Section \ref{sec:intdoping}  explores in more detail the intermediate ``momentum-space differentiated'' doping regime, studies the  momentum-selective regime and aspects associated with the pseudogap. Section \ref{sec:selective} then considers the sector-selective regime. In Sec.~\ref{sec:smallvslarge} we summarize our insights into the behavior of smaller size clusters. Finally, Sec.~\ref{sec:conclusion} is a summary and conclusion, also pointing out directions for future work. 

\section{Model and method \label{sec:overview}}
In conventional electronic-structure theory, band insulators are periodic crystals  in which all electronic bands are either filled or empty. A necessary condition for band-insulating behavior  is that the number of electrons per unit cell is even.  For the purpose of this paper we define a correlation-induced or  ``Mott'' insulator as a periodic crystal which has no broken symmetry but is insulating, even though the number of electrons per unit cell is not an even number.  The insulating behavior must arise from electronic correlations which are beyond the scope of conventional band theory.  The basic theoretical model used to study the correlation-induced transition is the one-orbital Hubbard model, consisting of electrons hopping among sites $i$ of a lattice and subject to an interaction $U$ which disfavors configurations with two electrons on a given lattice site. The model is most conveniently written in a mixed position-space/momentum-space representation as
\begin{equation}
H=\sum_{k\sigma}\varepsilon_kc^\dagger_{k\sigma}c_{k\sigma}+U\sum_i{\hat n}_{i\uparrow}{\hat n}_{i\downarrow}.
\label{Hdef}
\end{equation}
Here $c^\dagger_{k\sigma}$ is the Fourier transform of the operator $c^\dagger_{i\sigma}$ which creates an electron of spin $\sigma$ on site $i$ and ${\hat n}_{i\sigma}=c^\dagger_{i\sigma}c_{i\sigma}$ is the number operator for electrons of spin $\sigma$ on site $i$.  We specialize to a two-dimensional square lattice with hopping parameters such that the electron dispersion is 
\begin{equation}
\varepsilon_k=-2t\left(\cos k_x+\cos k_y\right)-4t{'}\cos k_x\cos k_y.
\label{epsilondef}
\end{equation}
We choose energy units such that $t=1$ and specialize to $t{'}=-0.15t$. A nonzero  $t{'}$ introduces a particle-hole asymmetry into the problem. The relative magnitude and sign of $t{'}$ are chosen  to be representative of the band structure of high temperature superconductors;\cite{Andersen94} we note that for very much larger amplitudes of $t{'}$ ($|t{'}|\gtrsim 0.3$) the behavior, especially on the electron-doped side, differs slightly from what is  discussed here.\cite{Gull09} We take $U=7t$; this value is slightly smaller than the $U\approx 9t$ believed to be relevant to high-temperature superconductors \cite{Comanac08} but is large enough that for all of the clusters we study a Mott insulating phase exists while being small enough to enable us to perform computations on large clusters with the resources available to us.

To solve the model we use the ``dynamical cluster'' (DCA) version of cluster dynamical mean-field theory.\cite{Hettler98,Maier05} In this approach the Brillouin zone is partitioned into $N$ equal-area tiles which we may label by $N$ central momenta $K_i$. The electron self-energy $\Sigma(k,\omega)$ is approximated by a piecewise constant function: $\Sigma(k,\omega)=\Sigma_{K_i}(\omega)$ if $k$ is contained within the tile labeled by $K_i$. The functions $\Sigma_{K_i}(\omega)$ are obtained from the solution of an auxiliary $N$-site quantum impurity model  with on-site interaction $U$ and  hopping and bath-coupling parameters determined by a self-consistency condition which may be written
\begin{equation}
G_K(i\omega_n)=\int^{K} d^2k\frac{1}{i\omega_n+\mu-\varepsilon_k-\Sigma_K(i\omega_n)}.
\label{Gdef}
\end{equation}
Here $G_K$ and $\Sigma_K$  are  Green's functions and self energies of the impurity model and $\int^{K}$ denotes the normalized integral over the momenta in the sector labeled by momentum $K$.    For details of the self-consistency procedure see Ref.~\onlinecite{Maier05}.

DCA schemes are distinguished by the number and arrangement of the tiles. Note that while (as far as is known) the method requires all of the tiles to be non overlapping and to  have a volume equal to $1/N$ of the Brillouin zone  both the {\em shape} of the tiles and their arrangement  may be freely chosen.  The freedom to adjust the tile shape was introduced  in recent $2$-site dynamical mean-field studies.\cite{Ferrero09,Ferrero09B} We restrict attention to tilings which respect the point symmetry of the lattice but consider different tiling shapes.  We use the $2$-site patching introduced in Ref \onlinecite{Ferrero09} and introduce an alternative patching, which we refer to as $4^*$, for the $4$-site cluster.  The specific clusters we consider  are shown in the five panels of   Fig. ~\ref{fig:clusters}.

Also shown as light lines in Fig.~\ref{fig:clusters} are the  Fermi surfaces of the noninteracting model for carrier concentrations corresponding to half filling and $10\%$, $20\%$, and $30\%$ hole doping.   The Fermi surface lines show  a deficiency of the standard $4$-site cluster:  for hole-dopings near to half filling the Fermi surface is almost entirely contained in the $(0,\pi)/(\pi,0)$ sectors; thus this cluster has a  difficult time capturing  momentum-space differentiation along the Fermi surface. The alternative $4^*$ patching shown in the middle panel of Fig.~\ref{fig:clusters} 
offers the possibility of capturing some of the zone-diagonal region of the Fermi surface within a different sector. 

The essential computational task is the solution of the quantum impurity model. To accomplish this we use continuous-time quantum Monte Carlo methods in the auxiliary field (CT-AUX) formulation \cite{Gull08_ctaux}. This method is an imaginary time method which yields the particle densities in each sector, along with the sector Green's function $G_K$ and the sector self-energy $\Sigma_K$.   From $G_K$ we obtain the sector occupancy $n_K$ via
\begin{equation}
n_K =G_K(\tau \rightarrow 0^-)= \frac{1}{\beta}\sum_n G_K(i\omega_n) e^{i\omega_n 0^+}.
\label{nKdef}
\end{equation}
Note that the sector Green's functions  are normalized in such a way that for $n_K=2$ all $k$ states in a sector are occupied by two electrons.  The total density is $n=2(\sum_K n_K)/N$ (the two is for spin).

In our analysis we work for the most part with sector quantities $G_K$, $\Sigma_K,$ and $n_K$.  We prefer to avoid the ``periodization'' or interpolation schemes which attempt to reconstruct continuous functions of momentum from the coarse-grained quantities which are the direct output of the calculation.   

\def\ImSigK{\Sigma''(K,0)}
\def\ImSigk{\Sigma''(k,0)}
Important quantities for the following discussion are the parameters  $\mu^*_k$, $Z_k,$ and $\ImSigk\equiv\Sigma''(k,\omega=0)$. These are defined generally for a Fermi liquid  in terms of the low-frequency limit of  the real $(\Sigma^{'})$ and imaginary $(\Sigma^{''})$ parts of the retarded electron self-energy $\Sigma(k,\omega)$   as
\begin{subequations}
\begin{align}
\Sigma(k,\omega) &\approx \mu^*_k-\mu+i \ImSigk  + \omega(1- Z_k^{-1})+....
\label{sigmalow}
\\
\mu^*_k& \equiv \mu + \Sigma'(k,0)
\label{mustardef}
\\
Z^{-1}_k& \equiv 1-
\left.\partial_{\omega} \Sigma'(k,\omega) \right|_{\omega=0} \ .
\label{zdef}
\end{align}
\end{subequations}
It will also be useful to consider 
\begin{equation}
\Gamma_k \equiv Z_k |\ImSigk|\ .
\label{gammadef}
\end{equation}
In the DCA approximation we use here these become piecewise constant functions of momentum; we denote the value appropriate to sector $K$ by suppressing the momentum argument and adding a subscript $K$. 

In the Fermi-liquid regime, these parameters express important aspects of electronic physics.  For completeness we briefly recall their meaning here. At low frequencies $\omega \rightarrow 0$,  the spectral function $A(k,\omega)=-\frac{1}{\pi}\text{Im} G(k,\omega)$ becomes
\begin{equation}
A(k,\omega)\approx \frac{1}{\pi}\frac{Z_k\Gamma_k}{\left(\omega+Z_k\left(\mu^*_k-\varepsilon_k\right)\right)^2+\Gamma_k^2}\ .
\label{GFL}
\end{equation}
$\mu^*_{k}$ determines the location of the renormalized Fermi surface (which is the locus of points $k_F$ for which $\varepsilon_{k_F}=\mu^*_{k_F}$). Thus a momentum dependence of $\mu^*_k$ signals a change in shape of the Fermi surface and more generally a shift in the mean energy of one momentum sector relative to the others. 
%
$\Gamma_k$ is the width of the quasiparticle peak. For a fixed  $k$ on the Fermi surface, $A(k_F,\omega)$ is peaked at $\omega=0$ and  the width in frequency is set by $\Gamma_k$.   A necessary condition for Fermi-liquid behavior is that $\Gamma_k$  be small, in which case $A(k,\omega)$ is characterized by a reasonably well-defined quasiparticle peak with frequency width given by $\Gamma_k$ and area given by the quasiparticle weight $Z_k$. The criterion $\Gamma_k<\pi T$ is the mathematical expression of the condition that the width of a thermally excited quasiparticle is less than its energy. In a Fermi liquid $\ImSigk\sim T^2$ as $T\rightarrow0 $ so $\Gamma$ is parametrically less than $T$. In this paper, however, since we cannot reach very low temperatures,  it will be useful to relax this definition and consider as a {\it quasi-Fermi liquid} any system where $(i)$ the Luttinger theorem is reasonably well obeyed, $(ii)$ $\Gamma_k$ decreases as $T$ decreases at all points along the renormalized Fermi surface, and $(iii)$ at all points along the Fermi surface $\Gamma_k<\pi T$.


Defining the bare velocity ${\vec v}_k=\partial \varepsilon_k/\partial {\vec k}$, the dispersion away from the Fermi surface is determined by the renormalized velocity
\begin{equation}
{\vec v}_k^*=Z_k\left(\frac{\partial \varepsilon_k}{\partial {\vec k}}+\frac{\partial \Sigma'(k,\omega=0)}{\partial {\vec k}}\right)=Z_K{\vec v}_k,
\label{vdef}
\end{equation}
where in the second equality we used the fact that in  the DCA approximation used here the self-energy is a piecewise constant function of momentum so that the  quasiparticle weight $Z_K$ coincides with the velocity renormalization. 

Note that dc transport properties  are controlled by the  mean free path
\begin{equation}
l_k=\Sigma^{''}(k,\omega)/\left|\partial_{\vec k}\varepsilon_k+\partial_{\vec k}\Sigma^{'}(k,\omega=0)\right|=\frac{\Sigma^{''}_K(\omega=0)}{v_k},
\label{ldef}
\end{equation}
i.e., by $\ImSigK$, not by $\Gamma_K$. We have used the piecewise continuity of the DCA self-energy in the second equality.

An estimate for the sector spectral function $A_K(\omega)$ in a range of order $T$ around the  Fermi level may be obtained from \cite{Trivedi95}
\begin{equation}
\beta G_K\left(\frac{\beta}{2}\right)=-\int \frac{d\omega}{2 T}\frac{A_K(\omega)}{\cosh\frac{\omega}{2T}},
\label{gtau}
\end{equation}
because the $\cosh^{-1}$ term implies that this integral is sharply peaked around $\omega=0$.

The quantities $\mu^*_{K}$, $Z_{K}$ and $\Sigma^{''}_K(\omega=0)$ are defined in terms of the real-axis self-energy while the impurity models are solved on the imaginary frequency axis. Because the studies  in this paper require the  survey of a wide range of clusters and dopings we have not been able to obtain data of sufficient quality to permit reliable analytical continuation of all of our cluster data. Different size clusters impose different computational burdens and (with the computational resources we have)  lead to data with differing statistical errors, introducing further uncertainty in the comparison of continuations based on data from differently sized clusters.  We therefore estimate $\mu^*_K$, $Z_{K}$ and the spectral function at zero energy $A_K(\omega=0)$ directly from the Matsubara-axis self energies.  $\mu^*_K$ is obtained by extrapolating the real part of  the Matsubara-axis self-energy to zero. $\Sigma^{''}_K(\omega=0)$ may be obtained as the extrapolation of the imaginary part of the Matsubara self-energy to  zero frequency. The   Kramers-Kronig relation connecting $\Sigma^{'}$ and $\Sigma^{''}$ implies that if $\Sigma''$ is small at low frequencies (as is the case in a Fermi liquid) $Z_{K}$ can be obtained from the extrapolation to zero frequency of the derivative of the Matsubara-axis $\Sigma(i\omega_n)$. We therefore estimate $\mu^*_{K}$, $Z_{K}$, and $\ImSigK$ by fitting the three lowest calculated Matsubara  frequencies to a quadratic form, which we use to extrapolate the value and  first derivative to zero Matsubara frequency.

This ``poor man's analytical continuation''  becomes less reliable at higher temperatures because the spacing between  Matsubara points is $2\pi T$ and therefore discretization errors become larger for higher $T$. At all temperatures we study we find that of the three quantities $\mu^*_{K}$ has the least uncertainties (in the Fermi-liquid regime) because the real part is continuous across the real-frequency axis and has a relatively weak temperature dependence.  $\ImSigK$ is less accurately determined than $\mu^*_{K}$ because it has a marked temperature dependence and becomes small at low $T$, and $Z_{K}$ is least accurately determined because it is a derivative.  

For momenta removed from the Fermi surface the interesting physics (in a Fermi liquid) requires knowledge of the self-energy at nonzero frequency, in particular at the frequency $\omega_{qp}$ which solves the ``quasiparticle equation'' $\omega_{qp}+ \Sigma'(k,\omega_{qp})-\varepsilon_k=0$. Our poor man's analytical continuation procedure does not provide much  information about the real-frequency self-energy at non-zero frequencies,  and the quantities $\mu^*_K$, $Z_K,$ and $\ImSigK$ do not have a clear meaning for states with $\varepsilon_k\neq \mu^*$. We therefore focus in the following on momentum  sectors which contain pieces of the Fermi surface and on regimes where reasonably long-lived quasiparticles exist. 

Further comparison of our methods  to analytic continuation is given in appendix \ref{appendix:continuations}.

\section{different doping regimes}
\label{sec:regimes}

In this section we characterize the different doping regimes discussed in the introduction in terms of the dependence of the   total electronic density $n$ on the chemical potential $\mu$ and in terms of the  partial occupancy $n_{K}$ of the patch associated with momentum $K$,  defined by Eq.~(\ref{nKdef}). 
\begin{figure}[t]
\includegraphics[width=0.90\columnwidth]{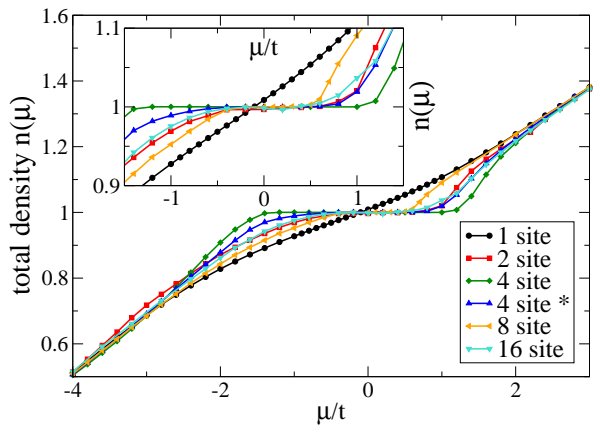}\\
\caption{\label{fig:nofmu}
Main panel: total electron density $n$ as a function of chemical potential $\mu$ for clusters considered in this paper at inverse temperature $\beta t=20$ for all clusters except $16$, where $\beta t =7.5$ is shown. Inset: expanded view of small chemical-potential region, highlighting region of Mott gap where $n=1$ independent of $\mu$.
}
\end{figure}

The total electron density $n$ is plotted in Fig.~\ref{fig:nofmu} as a function of the chemical potential $\mu$, for all clusters studied in this paper. For comparison, the single-site dynamical mean field (DMFT) result (i.e., the cluster with $N=1$) is also displayed.  Data are shown at the inverse  temperature $\beta =20/t$ for all clusters except $16$ sites, where we used $\beta =7.5/t$. We first observe that at half-filling ($n=1$) all $N>1$ cluster calculations yield an insulating state, revealed by the plateau in the $n(\mu)$ curve.    By contrast, single-site DMFT yields a metallic state [no plateau in $n(\mu)$], because the $U/t=7$ studied here  is smaller than the single-site DMFT value of the critical interaction strength $U\approx 12t$. The difference occurs because  cluster calculations take spatial correlations into account, and these stabilize the insulating state at smaller $U$ values than are needed in the single-site calculation.\cite{Jarrell01,Civelli05,Kyung06,Gull08_plaquette,Park08}

The range in chemical potential over which the $n(\mu)$ curve is flat can be used to define an estimate $\Delta_g$ for the insulating gap as $\Delta_g=\mu(n=1^+)-\mu(n=1^{-})$. Of course thermal effects mean that the $n(\mu)$ curve is not precisely flat. We adopt the criterion  that the chemical potential is within the gap if $0.99<n<1.01$.  Table~\ref{table:regimes}  presents results obtained for the different clusters, using this criterion, at  inverse temperature $\beta t = 20$ ($\beta t =7.5$ for $16$ site).  Except for the conventionally patched $4$-site cluster, all of the clusters studied give remarkably consistent estimates for $\Delta_g$.  We will explain below  why the conventionally patched $4$-site cluster is different.

We next turn to the doping dependence. It has previously been shown from small cluster studies  that at large enough doping, cluster corrections to the  single-site DMFT results  are small.\cite{Parcollet04,Civelli05} Figure \ref{fig:nofmu} shows that both on the hole-doped and electron-doped sides the differences between the results for different clusters become small  so that  the $n(\mu)$ curve is well described by the single-site DMFT result,  consistent with  previous findings. Below a critical chemical potential or critical density  the curves separate, indicating dependence on cluster geometry and therefore momentum differentiation. We define critical carrier concentrations $n^{h}_{\text{diff}}$ (on the hole-doping side)  and $n^{e}_{\text{diff}}$ (on the electron-doping side)  at which a significant momentum dependence appears in the electron self-energy. The $n_{\text{diff}}$ may  be estimated from the densities at which the $n(\mu)$ curves begin to separate, but as we shall see a better estimate may be obtained from analysis of  the sector dependence of the renormalized chemical potential $\mu^{*}_K$ as discussed in Sec.~\ref{sec:intdoping}. It is this latter analysis  which is used to obtain the densities shown in Table~\ref{table:regimes}.

\begin{table}
\begin{tabular}{|c||l|l|l|l|l|}
\hline
Cluster size & $n_{\rm{diff}}^h$ & $n_{\rm{diff}}^e$ & $\Delta_g$ & $n_{\rm{SST}}^h$ & $\Delta_{\rm{SST}}$ \\
\hline\hline
2& 0.66 & 1.27 & 1.4 & \ \ -   & \ \ -\\
4& 0.65 & 1.38 & 2.6 & \ \ -   & 2.6\\
$4^*$& 0.69 & 1.39 & 1.8 & 0.96   & 2.4\\
8& 0.72 & 1.23 & 1.1 & 0.93   & 1.9\\
16& 0.65 & 1.35 & 1.4 & 0.91   & 2.1\\\hline
\end{tabular}
\caption{Characteristic densities for onset of momentum differentiation on the hole-doped ($n^{h}_{\text{diff}}$) and electron-doped sides ($n_{\rm{diff}}^e$), estimated gap of the Mott insulator ($\Delta_g$), location of the sector-selective transition on the hole-doped ($n_{\rm{SST}}^h$) side (there is no SST transition on the electron-doped side for the parameters chosen in this paper), 
and gap of the sector-selective regime ($\Delta_{\rm{SST}}$).
}
\label{table:regimes}
\end{table}

\begin{figure*}[bth]
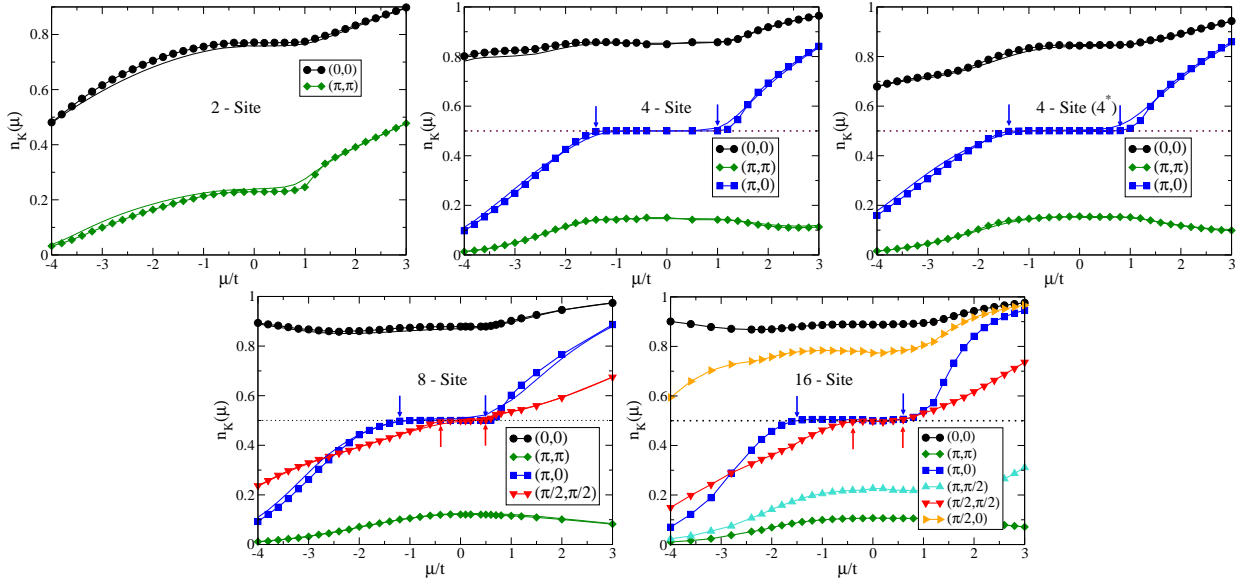

\includegraphics[width=0.3\textwidth]{./n_beta_2site.eps}
\includegraphics[width=0.3\textwidth]{./n_beta_4site.eps}
\includegraphics[width=0.3\textwidth]{./n_beta_4site_improved.eps}\\
\includegraphics[width=0.3\textwidth]{./n_beta_8site.eps}
\includegraphics[width=0.3\textwidth]{./n_beta_16site.eps}
\caption{\label{fig:partialnofmu}
Partial occupancies $n_K$ in the different patches of the Brillouin zone, as a
function of chemical potential $\mu$, for different clusters.
Inverse temperature is $\beta t=20$, thin lines and $16$-site cluster: $\beta t=7.5$.
The arrows indicate the points where
the sector densities deviate from the plateau values; the chemical-potential difference between the arrows defines the
range of chemical potential corresponding to the Mott insulator ($\Delta_g$)
and to the sector-selective regime ($\Delta_{\rm{SST}}$). The Mott insulating density $n=1/2$ is shown as the dotted line.
}
\end{figure*}
\begin{figure*}[tbh]
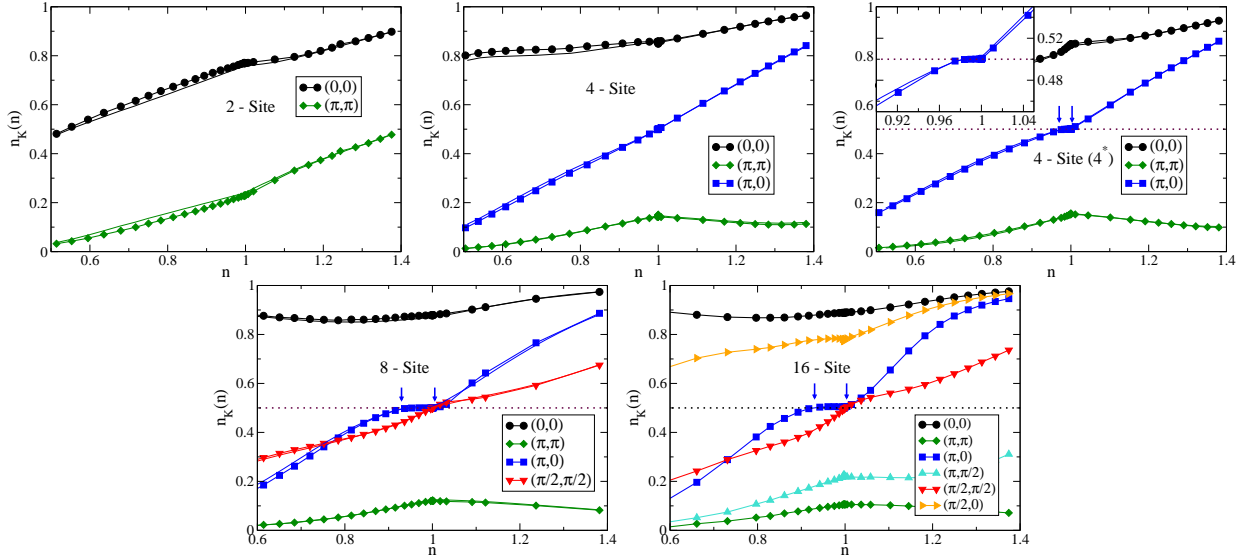

\includegraphics[width=0.3\textwidth]{./n_beta_n_2site.eps}
\includegraphics[width=0.3\textwidth]{./n_beta_n_4site.eps}
\includegraphics[width=0.3\textwidth]{./n_beta_n_4site_improved.eps}\\
\includegraphics[width=0.3\textwidth]{./n_beta_n_8site.eps}
\includegraphics[width=0.3\textwidth]{./n_beta_n_16site.eps}
\caption{\label{fig:partialnofn}
Partial occupancies $n_K$ in the different patches of the Brillouin zone, as a function of total electron density $n$ for the different clusters.
Inverse temperature $\beta t=20$, thin lines and $16$-site cluster: $\beta t=7.5$.
}
\end{figure*}

Figures \ref{fig:partialnofmu} and \ref{fig:partialnofn} display the occupancies $n_K$ of the different sectors, respectively, as a function of  the chemical potential $\mu$ and the total density $n$.  For the clusters with $N<16$ these figures display data for two temperatures: $\beta t=20$ (symbols and heavy lines) and $\beta t=7.5$ (light lines), showing that the temperature dependence is weak. 

Figure \ref{fig:partialnofmu} shows that for all clusters except the $2$ site, there is  a range of chemical potentials where the density in the $(0,\pi)$ sector remains locked at $n_{(0,\pi)}=1/2$ corresponding to a half-filled patch. The chemical-potential range over which the density in the $(0,\pi)$ sector is flat defines a gap, $\Delta_{\rm{SST}}$,  which is also shown in Table~\ref{table:regimes} and is of the same order for all clusters.   The presence of a gap is confirmed by the density of states estimates discussed in Sec.~\ref{sec:selective}.  

The regime of chemical potentials where the $(0,\pi)$ sector is incompressible while other sectors can be doped is a signature of the sector-selective regime discussed in previous work.\cite{Werner09,Gull09} We see that the phenomenon
is robust, occurring in $4^*$-, $8$-, and $16$- site clusters. It is also interesting to note that while (for the parameters we studied) the conventionally patched $4$-site cluster does not have a sector-selective regime, the change in momentum-space patching involved in going from $4$ to $4^*$ gives the cluster the possibility of distinguishing nodal and antinodal excitations, and  the transition reappears. This, along with the similar behavior of the $8$- and $16$-site clusters, is strong evidence  that the sector-selective transition (SST) is a generic phenomenon that appears wherever the momentum patching allows it. It  reveals strong momentum differentiation and the formation of an antinodal pseudogap at low hole-doping.

A different perspective on the $(0,\pi)$ sector gap is provided by the plots as a function of the total density shown in  Fig.~\ref{fig:partialnofn}. We see that in the $4^*$-, $8$- and $16$-site clusters  the partial density in the $(0,\pi)$ sector remains locked at $n_{(0,\pi)}=1/2$ for a range of densities while the density in other sectors changes. This behavior,  that for some range of doping the antinodal patch remains incompressible while the other patches accommodate the dopants,  is observed only on the hole-doped side (for the parameters $U/t,t^\prime/t$ studied here) and defines the sector-selective regime.   The boundary, $n=n_{\rm{SST}}^h$, of the density range over which the $(0,\pi)$ sector is incompressible   is a true $T=0$ phase transition of the self-consistent DCA equations associated with a specific cluster. For concreteness we define  $n_{\rm{SST}}^h$ as the value of the total density $n$, where $|n_{(0,\pi)} - 0.5| = 0.005$ and we report these values in Table~\ref{table:regimes}.  

Unlike the  $4^{*}$-, $8$- and $16$-site clusters, the conventionally patched $4$-site cluster does not display a sector-selective
regime. However, we see from Table~\ref{table:regimes} that the insulating gap for the conventionally patched $4$-site cluster has a magnitude very similar to the $(0,\pi)$ sector gap found in the $4^*$-, $8$- and $16$-site clusters.  This makes the origin of the difficulties of the $4$-site cluster clear: in this cluster the $(0,\pi)$  patch   must fill two roles. First, this patch contains the $(0,\pi)$ point so that its behavior must represent the physics of the pseudogap. Second, the patch contains essentially all of the Fermi surface, and therefore must represent the behavior of the gapless nodal quasiparticles.  The modified $4^*$ patching avoids this problem because the $(0,0)$ sector contains the nodal portion of the Fermi surface so that in this cluster the $(0,\pi)$ sector does not have two roles.  We also note that the critical density for the onset of the sector-selective regime is about $0.92$ for the $8$- and $16$-site clusters, but closer to $n=1$ for the $4^*$ cluster. We believe the difference arises because, while the $4^*$ cluster does allow some nodal/antinodal differentiation, the portion of noninteracting Fermi surface which is outside of the $(0,\pi)$ sector remains comparatively small.

The $2$-site cluster is also different: neither of the two momentum sectors exhibits a plateau outside of the insulating regime although previous studies of the two-patch cluster using a larger $t'$ (Refs.~\onlinecite{Ferrero09} and \onlinecite{Ferrero09B}) have revealed a sector-selective regime for a range of hole-dopings. We shall see in Sec.~\ref{sec:selective} that the difference arises from a difference in the nature of the sector-selective transition in this cluster.  

\begin{figure}[t]
\includegraphics[width=0.9\columnwidth]{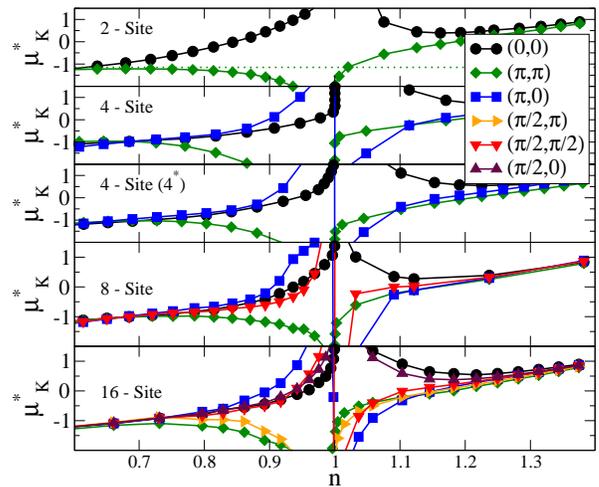}\\
\caption{\label{fig:allclustersmueff}
(Color online) Renormalized chemical potential $\mu^*_K = \mu -\Sigma'(K,\omega=0)$ for
different momentum sectors, plotted vs. particle number $n$  in (from top to
bottom panels) $2$-, $4$-,$4^*$-, $8$- and $16$- site clusters. Dotted line
(green): location of lower band edge for sector $(\pi,\pi)$ in the
$2$-site cluster.
}
\end{figure}

%

\section{momentum-space differentiation \label{sec:intdoping}}

In this section we analyze in more detail the momentum-space differentiated regime introduced in the previous section.
We begin our discussion with  $\mu^*_K$, Eq.~(\ref{mustardef}), shown as a function of total density $n$ in Fig.~\ref{fig:allclustersmueff}. For sufficiently high levels of electron or hole doping, $\mu^*_{K}$ becomes almost independent of momentum sector within a cluster and indeed takes the same value independent of cluster. As the doping is decreased toward half filling the traces separate, indicating that the self-energy begins to depend on  momentum. The onset of momentum-space differentiation does not (to our knowledge) correspond to a phase transition. Indeed at any carrier concentration the self-energy has some momentum dependence: perhaps weak but, in general, nonzero so that  any definition of  the doping at which the variation with momentum becomes significant involves some arbitrariness. In Table~\ref{table:regimes}, we estimate this density  $n^{h}_{\text{diff}}$ (on the hole-doping side)  and $n^{e}_{\text{diff}}$ (on the electron-doping side) as the density where  $\max_{K,K'}|\mu^{*}_{K} - \mu^{*}_{K'}| > 0.2$. 

\begin{figure}[b]
  \includegraphics[width=0.45\columnwidth]{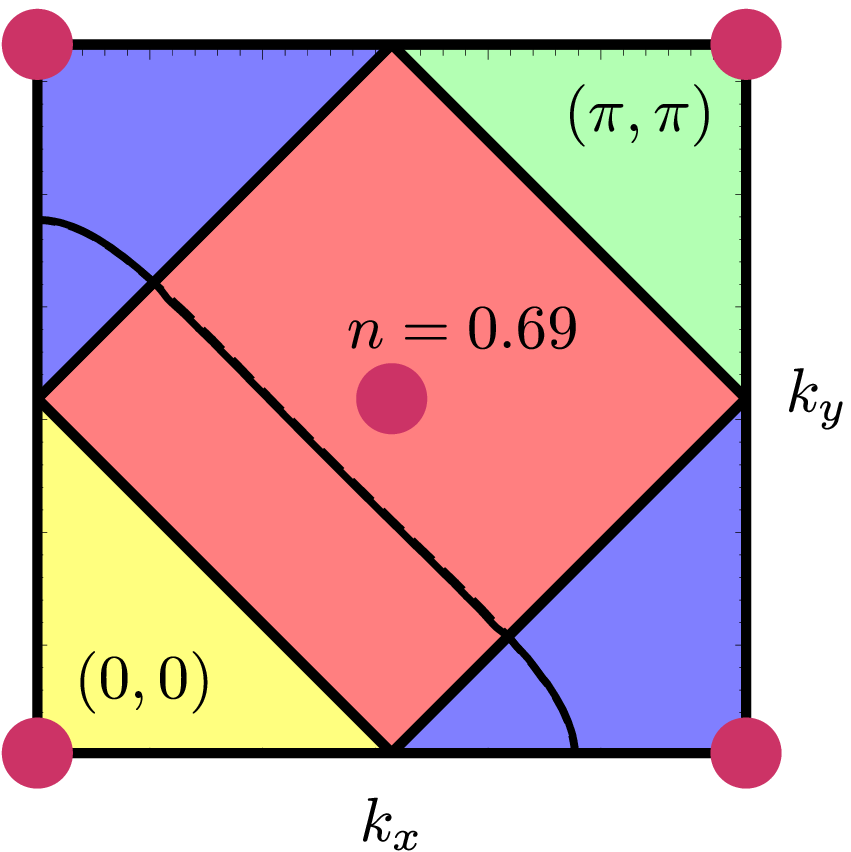}
  \includegraphics[width=0.45\columnwidth]{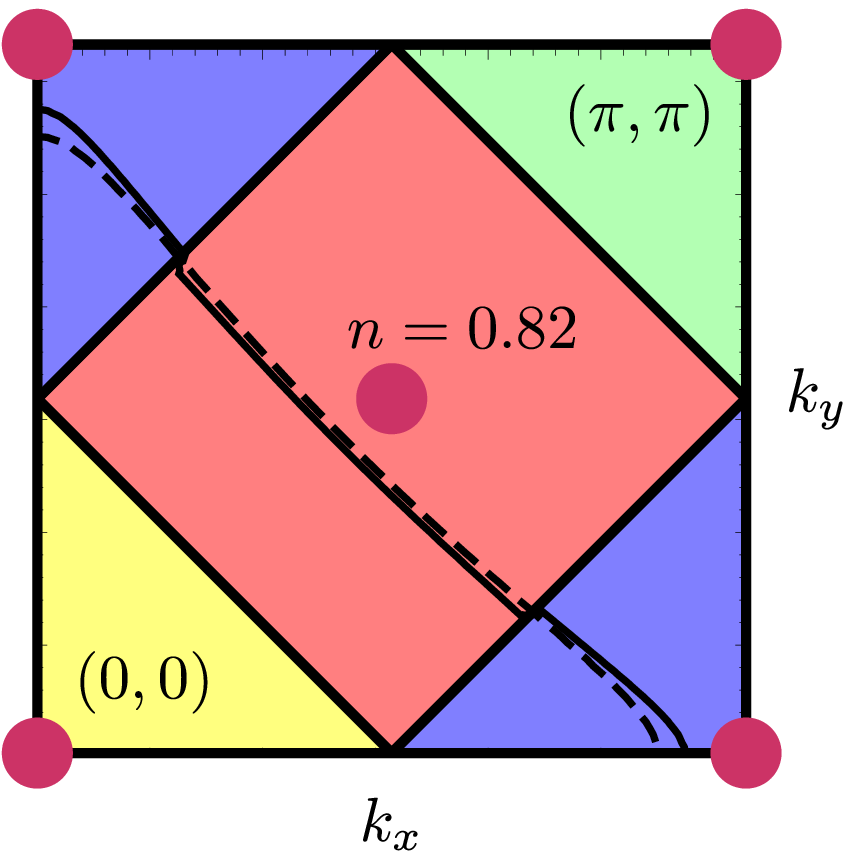}\\
\caption{\label{fig:brokenfs} Plot of the upper quadrant of the Brillouin zone showing the Fermi surface as a black line.
The dashed line shows the noninteracting Fermi surface with the same total density.
The sectors of the $8$-site cluster are also displayed (see Fig.~\ref{fig:clusters}). Left panel: $n\approx 0.69$ (in isotropic Fermi-liquid regime);
right panel $n\approx 0.82$ (in momentum-space-differentiated regime). All Fermi surfaces are consistent with the Luttinger theorem.}
\end{figure}

A momentum dependence of $\mu^{*}$ implies a change in shape of the Fermi surface. The two panels of Fig.~\ref{fig:brokenfs} show the Fermi surface computed for the $8$-site cluster in one Brillouin zone quadrant at two dopings, one (left panel) in the high-doping isotropic Fermi-liquid regime and one (right panel) in the intermediate doping momentum-space differentiated regime. In the isotropic Fermi-liquid regime the very weak $k$ dependence of $\Sigma$ means that the Fermi surface of the interacting model coincides with the noninteracting Fermi surface (at the same density). However, in the momentum-space differentiated regime the shape changes.  The piecewise constant nature of the DCA self-energy means that  the DCA Fermi surface is also a piecewise constant approximation to the true, renormalized Fermi surface.  We see that the interaction has the effect of enhancing the Fermi-surface curvature  so that the renormalized Fermi surface corresponds, in effect, to a quasiparticle band structure with a bigger $t{'}$ than the bare theory, consistent with the suggestions of Refs.~\onlinecite{Maier02} and \onlinecite{Maier03}.

An important result of Fermi-liquid theory is the ``Luttinger theorem,''\cite{Luttinger60} which states that the volume of $k$ space contained within the Fermi surface, defined by $\varepsilon_k+\text{Re}[\Sigma(k,\omega=0)]=\mu$, is equal to the density of particles per unit cell $(\text{mod} 2)$. We have verified that both Fermi surfaces are consistent with the Luttinger theorem to within $1\%$. Corresponding plots made for dopings outside the Fermi-liquid regime are not consistent with the Luttinger theorem because the ``Fermi surface,'' although mathematically defined, is not physically meaningful. We also note that the proof of the Luttinger theorem relies in an essential way on an integration over the whole zone; it is not obeyed sector by sector, meaning that the $n_K$ defined from the sector Green's function need not correspond to the sector area contained within the Fermi surface. 

\begin{figure}[b]
\includegraphics[width=0.9\columnwidth]{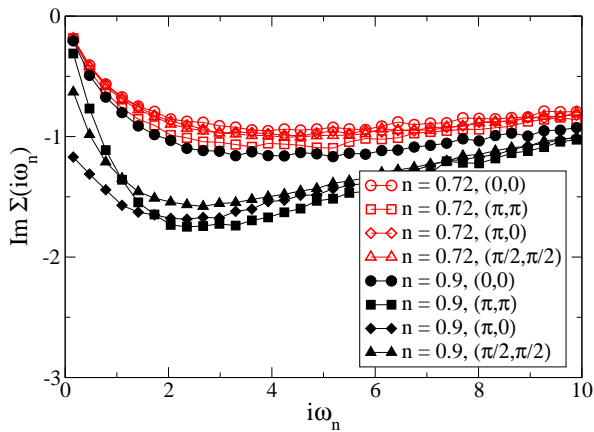}
\caption{\label{fig:selfenergy_8site} Imaginary part of Matsubara self-energies 
of the $8$-site cluster for two densities at inverse temperature $\beta t=20$, for various sectors.
}
\end{figure}

It is worth emphasizing that the isotropic Fermi-liquid regime, in which single-site DMFT  is barely corrected by cluster expansions, is not a trivial weak-coupling regime. This may be seen from  Fig.~\ref{fig:selfenergy_8site}, which shows the full frequency dependence of the imaginary parts   of all components of the self-energy of the $8$-site cluster  for dopings $n=0.72$ (in the isotropic Fermi-liquid regime) and $n=0.9$ (momentum-differentiated regime). In the isotropic regime the relative differences in self-energy between different sectors are small whereas they are large in the differentiated regime. 

\begin{figure}[t]
\includegraphics[width=0.9\columnwidth]{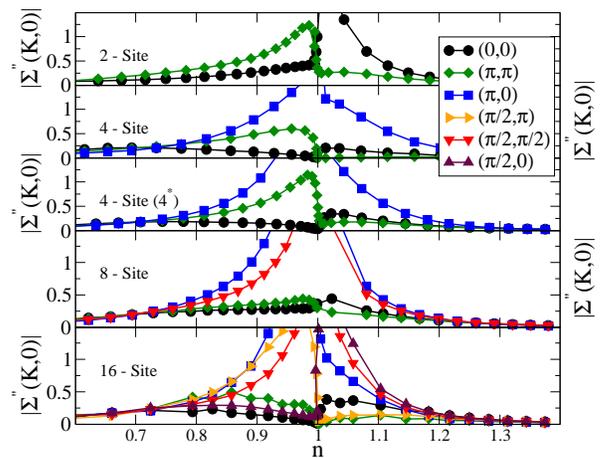}
\caption{\label{fig:allclusterssigma0} $|\ImSigK|$  for different momentum sectors plotted vs particle number $n$  in (from top to bottom panels) $2$-, $4$-, $4^*$-, $8$-, and $16$- site clusters.  }
\end{figure}

We next turn to  the scattering rate for low-energy excitations, which we estimate from   $\ImSigK$, the extrapolation to $i\omega_n=0$ of the imaginary part of the Matsubara self-energy [Eq.~(\ref{sigmalow})]. Figure \ref{fig:allclusterssigma0} presents results obtained  at the relatively high temperature $\beta t=7.5$ where we have data for all clusters and dopings. We see that at high dopings the scattering rates are small and (within our numerical accuracy) independent of cluster size and sector, reflecting the reasonably isotropic nature of the high-doping regimes. As doping is decreased the curves clearly exhibit an onset of momentum-space differentiation at a  doping $\sim 0.25$ consistent within errors with the doping estimated from $\mu^{*}$ and  again only weakly dependent on cluster size. 

\def\ImSig#1{\Sigma''\bigl (#1,0 \bigr)}
Momentum-space differentiation is marked by a strong relative increase with decreasing doping of $\ImSig{(0,\pi)}$  and  a less strong increase in  $\ImSig{(\pi/2,\pi/2)}$ (for the clusters which provide access to this sector).  These are the two sectors which contain the Fermi surface and for which $\ImSigK$ has meaning as a scattering rate. Comparison of electron and hole dopings shows  that while momentum-space differentiation sets in at about the same absolute value of doping in the two cases, the degree of differentiation between  sectors $(0,\pi)$ and $(\pi/2,\pi/2)$ is greater on the hole-doped side than on the electron-doped side.

\begin{figure}[b]
\includegraphics[width=0.99\columnwidth]{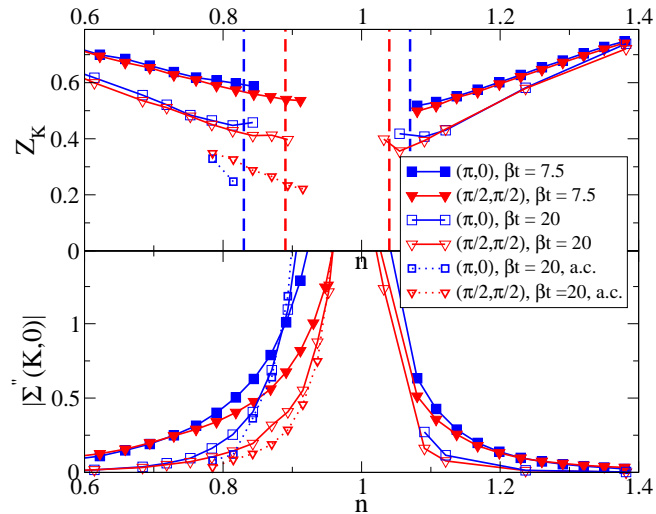}
\caption{\label{fig:sigma0beta15258site} Estimators of quasiparticle weight/velocity renormalization factor $Z$ (upper panel) and scattering rate $|\ImSigK|$ obtained from extrapolation of  Matsubara-axis self energies for the $K=(\pi,0)$ and $K=(\pi/2,\pi/2)$ sectors of $8$-site cluster at temperatures $T=t/7.5$ and $t/20$, along with estimates obtained by maximum entropy analytical continuation (a.c.). Vertical dashed lines denote the boundaries separating the regime consistent with Fermi-liquid theory ($\beta \Gamma_{K} < \pi$) from the regime that is clearly not Fermi liquid.
}
\end{figure}

We turn now to a more detailed examination of results from  the $8$-site cluster, which is large enough allow a direct comparison of   the nodal and antinodal regions of the Fermi surface, but is small enough to allow detailed computations down to relatively low temperatures.   The two panels of Fig.~\ref{fig:sigma0beta15258site} show $\ImSigK$ and the quasiparticle weight/velocity renormalization $Z_{K}$ for the nodal $K=(\pi/2,\pi/2)$ and antinodal $K=(\pi,0)$ sectors as a function of doping at a relatively low and a relatively high temperature.

Comparison of the two panels of Fig.~\ref{fig:sigma0beta15258site} shows that the momentum-space differentiation is marked primarily by a variation in scattering rate. As doping is reduced, the Fermi-surface scattering rates increase rapidly and a marked difference between the two Fermi surface sectors develops with the antinodal sector $K=(\pi,0)$ characterized by a much more rapidly growing scattering rate. Further, the scattering rates exhibit a pronounced particle-hole asymmetry.  However, while the  inverse  mass enhancement/velocity renormalization $Z_{K}$ decreases as doping is decreased, the variation with doping is much less dramatic and, interestingly, there is very little particle-hole asymmetry or difference between the two momentum sectors.   We also note that the nodal quasiparticle residue $Z_{(\pi/2,\pi/2)}$ appears to  extrapolate to a non-zero value at $n=1$. (A different result was found using self-energy interpolations in superconducting state cellular dynamical mean field (CDMFT) calculations on $4$-site clusters.\cite{Civelli08,Haule07plaquette}) This is inconsistent with the Brinkman-Rice theory but qualitatively consistent with data on high-$T_c$ materials, where photoemission measurements indicate a zone-diagonal quasiparticle velocity which is only weakly doping dependent.\cite{Shen03} (Very recent measurements indicate that if the velocity is measured on very low scales, below the resolution of the numerics in this paper or of previous photoemission data a stronger doping dependence of the velocity is found.\cite{Vishik10})

\begin{figure}[t]
\includegraphics[width=0.95\columnwidth]{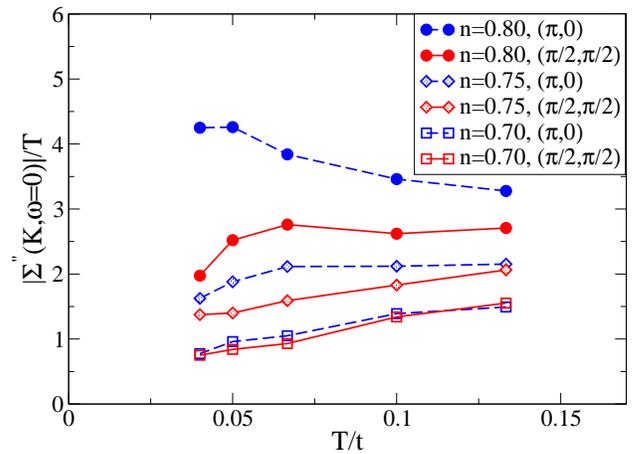}
\caption{\label{fig:sig2oft.eps} (Color online) Temperature dependence of  $|\ImSigK|$ for $(\pi,0)$ sector (dashed lines, black) and $(\pi/2,\pi/2)$ sector (solid lines, red) of $8$-site cluster at doping $0.7$ (open symbols), $0.75$ (light symbols), and $0.8$ (filled symbols).
}
\end{figure}

Figure \ref{fig:sig2oft.eps} presents the temperature dependence of the  nodal and antinodal scattering rates obtained for the $8$-site cluster for selected densities. To highlight the temperature dependence we plot $\Sigma^{''}/T$. While the temperature range accessible to us is too limited to establish any specific form of temperature dependence it is clear that at the higher doping isotropic Fermi-liquid regime ($n=0.70$),  the scattering rates drop faster than linearly at low $T$  while at the lower doping ($n=0.80$; momentum-differentiation regime)  the two sectors have different temperature dependence at low temperature with the nodal sector scattering rate vanishing more rapidly than $T$ at low $T$ and the antinodal rate vanishing less rapidly. At the intermediate doping $n=0.75$ on the boundary between the two regimes the behavior is intermediate.  These features are in qualitative agreement with the momentum-space variation in  the electronic mean free path inferred from angular-dependent magnetoresistance experiments.\cite{Abdel-Jawad07,French09} For highly overdoped cuprates these experiments reveal  a scattering rate which is reasonably isotropic around the Fermi surface and exhibits  a relatively conventional temperature dependence. Below a critical doping a momentum-space differentiation appears with the antinodal scattering rate being larger and exhibiting a weaker temperature dependence.

\section{Sector Selective Regime \label{sec:selective}}

In this section we discuss in more detail the sector-selective regime, which has been characterized in Sec.~\ref{sec:regimes} by the existence of a plateau in the $n_K(\mu)$ curve for the sectors containing $(0,\pi)$ (and symmetry-related points)  and the absence of such a plateau in the sectors containing $(\pm\pi/2,\pm\pi/2)$.   We first observe that, as seen, for example, in Fig.~\ref{fig:allclusterssigma0}, the approach to the sector-selective phase in clusters $4^*$, $8,$ and $16$ is  marked by a rapid increase in the absolute value of $\ImSig{(0,\pi)}$ and $\ImSig{(\pi/2,\pi/2)}$, with  $\ImSig{(0,\pi)}$ being much larger. In the $(0,\pi)$ sector the increase is associated with the formation of a pole in the sector self-energy. This pole is responsible for the sector-selective insulating behavior,  as discussed in detail in Refs.~\onlinecite{Werner09,Gull09,Lin10}. However, we also see that as doping is decreased the $(\pi/2,\pi/2)$ sector self-energy increases, and in fact becomes large enough that (at least at the temperatures accessible to us in this study) this sector is clearly in a non-Fermi-liquid regime (although as seen it is compressible). It is an interesting, but so far unresolved, question whether the nodal sector is intrinsically non-Fermi liquid in this doping regime or whether the Fermi temperature is simply lower than the lowest temperatures we can access.

\begin{figure*}[t]
 \includegraphics[width=0.95\textwidth]{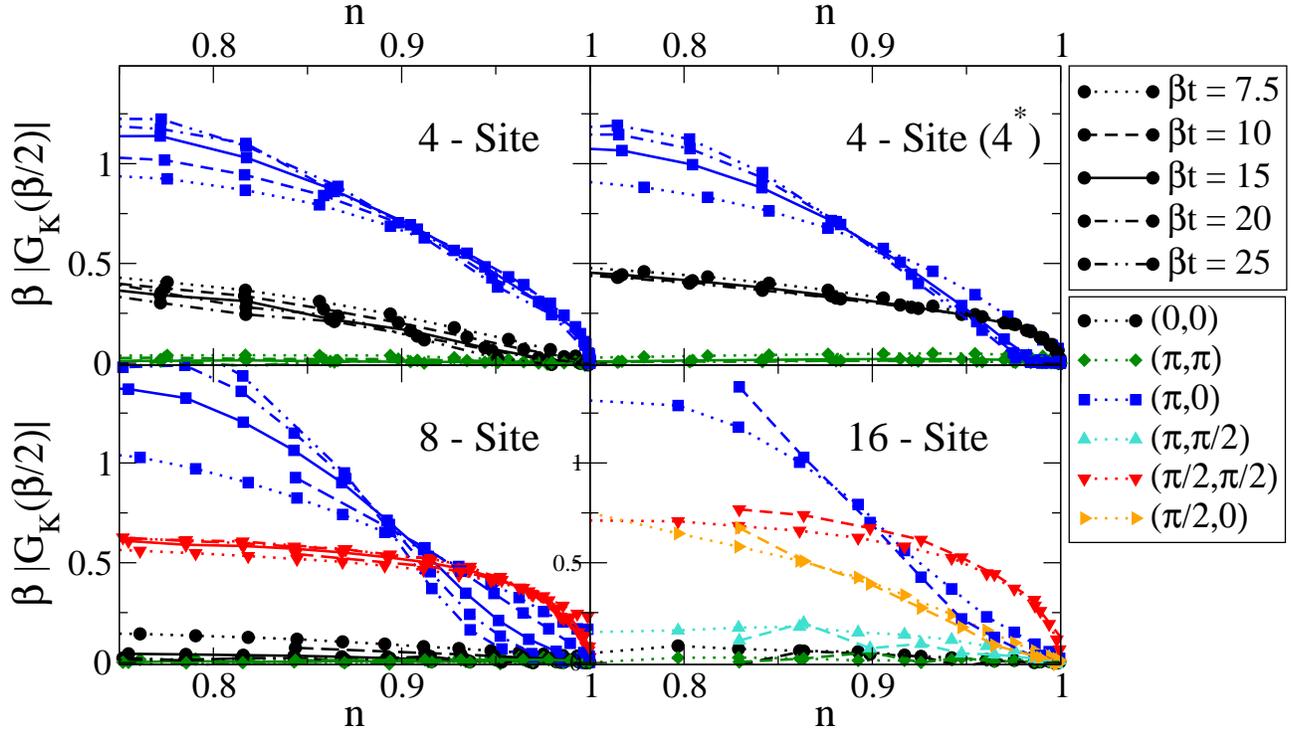}
 \caption{\label{fig:estimator_density_n} Estimator for the sector density of states, $\beta |G_K(\beta/2)|$,
as a function of density (hole-doped case only), for temperatures ranging from $\beta t=7.5$ to $\beta t=25$,
and for the clusters  $4$, $4^{*}$, $8$, and $16$.
Symbol styles denote patches, 
line styles denote different temperatures.
A clear crossing point as a function of temperature is visible in sector $(\pi,0)$, for 
the clusters $4^*$, $8$, and $16$, but is absent for the standard $4$-site cluster.
}
\end{figure*}

We next show that, as expected on physical grounds, the plateau in sector density $n_K$ coexists with a gap in the sector density of states. Figure \ref{fig:estimator_density_n} shows  $\beta |G_K(\beta/2)|$  [Eq.~\ref{gtau}] which, as $T\rightarrow 0$, converges to the sector density of states at the Fermi level, $A_K(\omega=0)$. We first focus on the lower left panel, showing results for the $8$-site cluster at several temperatures. We see, as previously shown,\cite{Werner09,Gull09} that the $(0,0)$ and $(\pi,\pi)$ sectors have essentially no low-energy density of states. This is expected because the Fermi
surface does not pass through these points at any of the densities displayed. The $(\pi/2,\pi/2)$ sector shows a density of states which at higher doping, in fact, takes the Fermi-liquid value, but in the sector-selective regime is suppressed from the
Fermi-liquid value [consistent with the large value  in this regime of the scattering rate in the  $(\pi/2,\pi/2)$ sector] but at lower temperatures is weakly temperature dependent. The weak temperature dependence suggests, but with our limited temperature range does not prove, that the sector is in a non-Fermi-liquid regime. By contrast the sector $(0,\pi)$ results display a strong temperature dependence: for carrier concentrations  above  $\sim 0.9$ the sector density of states is a strongly decreasing function as temperature is decreased, indicating that a gap is forming in this sector. 

Now consider the other panels in the plot, beginning with the $16$-site cluster. The temperature range accessible to us  is
more limited, because at  high doping our ability to calculate at  low temperature is  hampered by a sign problem. However, within the available resolution we see that  the  $16$-site cluster behaves in the same manner as the $8$-site cluster. This is in contrast to the conventionally patched $4$-site cluster, where we see that the density of states in the $(0,\pi)$ sector, while suppressed and weakly temperature dependent, only vanishes at $n=1$, similar to the behavior of the $(\pi/2,\pi/2)$ sector in the $8$- and $16$-site clusters. The $(0,0)$ sector is simply gapped, similar to the $(0,0)$ sectors in the larger clusters.    The alternative patching of the $4^{*}$ cluster  remedies  this behavior by including some of the noninteracting Fermi surface in the $(0,0)$ sector, and we see that this cluster behaves more similarly to the $8$- and $16$-site clusters.  In the $4^{*}$ cluster the sector-selective transition exists but is seen at a lower doping than in the larger clusters [corresponding to the smaller part of the Fermi surface that $(0,0)$ intersects], and the behavior of the $(0,0)$ sector is intermediate between that of a nodal sector in a larger cluster and the conventionally patched $4$-site cluster. 

\section {Small versus large clusters} \label{sec:smallvslarge}

In this section we  compare the results obtained on small ($2$- and $4$-site) clusters to those obtained on the larger $8$- and $16$-site clusters. Roughly speaking, the sector quantities $G_K$ and $\Sigma_K$ correspond to averages of the intrinsic quantities $G(k,\omega)$ and $\Sigma(k,\omega)$ over appropriate regions of the Brillouin zone. As the cluster size is reduced, the averages extend over wider regions of momentum-space.  While it is clear that very small clusters will only be able to capture a very coarse-grained version of momentum-dependent physics, it is nevertheless interesting to understand the strengths and weaknesses of small cluster studies, both because it is probably not feasible to use larger clusters to study multiorbital models and because even for the one orbital Hubbard model much more accurate data can be obtained, down to much lower temperatures, on small clusters than on large clusters. 

We begin by reconsidering the cluster size dependence of the scattering rates shown in  Fig.~\ref{fig:allclusterssigma0}, starting from the $16$-site cluster down to the smallest clusters.  The  self energies obtained on the $16$- and $8$-site clusters for the $K=(0,0),$ $(\pi,\pi),$ $(0,\pi)$, and $(\pi/2,\pi/2)$ patches are very similar. In both of these clusters the different patches correspond to regions of the zone with different physical properties.  When the momentum resolution starts to be reduced, the patches become bigger and the cluster quantities involve averages over regimes with different physical properties and therefore yield results which are in some sense intermediate.  
For example, 
the $(\pi,0)$ sector of the $4$-site cluster is seen to have  an intermediate behavior between the $(\pi,0)$ and $(\pi/2,\pi/2)$ sectors  of the $8$- or $16$-site clusters while the $(\pi,\pi)$ sector of the $4^{*}$  cluster has behavior somewhat closer to that of the $(\pi,0)$ sector of the larger cluster. In general, it appears that  if this basic fact is borne in mind as results are interpreted, the smaller clusters do remarkably well in capturing the important qualitative behavior, although the clear quantitative differences call into question the validity of the various interpolation schemes.  The example of the $4$ and $4^{*}$ clusters also shows how a physically motivated choice of patching (or even a comparison between different patching choices) can reveal important physics, in this case better separating the nodal and antinodal sectors.


Let us comment in more detail on the specific case of the 2-site cluster, which was analyzed in great detail in  Refs.~\onlinecite{Ferrero09} and \onlinecite{Ferrero09B}. In these papers the second neighbor hopping $t'=-0.3t$ was used   and a doping-driven  sector-selective transition was found, but of a different character than the sector-selective transitions discussed here and in Refs.~\onlinecite{Werner09} and \onlinecite{Gull09}.   The transition found in Refs.~\onlinecite{Ferrero09} and \onlinecite{Ferrero09B} is 
caused by a renormalization of the effective low-energy chemical potential by the  interactions and can be understood as follows. For a given DCA patching of the Brillouin zone, a necessary condition for the Fermi surface of the interacting system to penetrate within the momentum-$K$ patch of the Brillouin zone is that the renormalized chemical potential $\mu^*_K$ satisfies the condition $\mu^*_K\in[\epsilon_{\rm{min}}^K,\epsilon_{\rm{max}}^K]$, where   $\epsilon_{\rm{min},\rm{max}}^K$ are the minimum/maximum energy spanned by the noninteracting band $\epsilon_k$ as $k$ is varied within patch $K$. This relation directly follows from Eq.~(\ref{Gdef}) relating the coarse-grained Green's function $G_K$ to the self-energy $\Sigma_K$. 

For  very  high hole doping (low carrier concentration) the Fermi surface is contained entirely within the $(0,0)$ patch of the $2$-site cluster. As the doping is decreased, the Fermi surface expands and in the noninteracting model will penetrate the outer patch at some critical doping, which depends on $t'$. However, in the momentum-differentiated regime the sector dependence of the renormalized chemical potential in the $2$-site cluster  is such as to push the renormalized Fermi level away from the outer patch [in contrast to the larger clusters, where the $\mu^{*}_K$ is such as to push the Fermi surface in the $(0,\pi)$ patch closer to half filling]. This effect gets larger as half filling is approached. For the $t'$ and $U$  studied  in Refs.~\onlinecite{Ferrero09} and \onlinecite{Ferrero09B}, for some range of dopings the Fermi level was able to penetrate into the outer patch, but as doping was further reduced the increase in  $\mu^{*}_K$ for the outer patch was rapid enough to push the Fermi level out again, resulting in a sector-selective transition of the band-insulator type.  For the smaller $t^{'}$ studied here  the increase in $\mu^{*}_K$ for the outer patch is such as to prevent the Fermi level from ever reaching the outer patch. We may therefore regard the $2$-site cluster as being in a sector-selective regime  for all hole dopings, where the antinodal states are gapped and suppressed compared to the non-interacting system. This physics is particular to the $2$-site cluster, but as noted in Refs.~\onlinecite{Ferrero09} and \onlinecite{Ferrero09B} does capture (as best this cluster can) the momentum-space differentiation physics which is important for cuprates. 

\section{Conclusion \label{sec:conclusion}}

This paper has  presented ``DCA'' cluster dynamical mean-field calculations based on  $2$-, $4$-, $8$- and $16$- site clusters of the two-dimensional Hubbard model, with hopping parameters appropriate  to high-$T_c$ materials and an interaction which is slightly weaker than in the actual compounds, but strong enough to display the essential physics. The comparison of results obtained on different sized clusters shows unambiguously that in the two-dimensional Hubbard model with underlying particle-hole asymmetry appropriate to high-$T_c$ materials, the doping-driven metal-insulator transition proceeds through several intermediate regimes, but  is different for electron-doped than for hole-doped materials. For both hole-doped and electron-doped materials a high doping isotropic Fermi-liquid regime gives way as doping is reduced to a momentum-space-differentiated Fermi-liquid regime, in which the quasiparticle residue (equivalent to the velocity renormalization in the approximation used here) depends only weakly on momentum but the electron-scattering rate acquires a strong dependence, being small for states near the zone-diagonal and large for states near the zone face.  The momentum dependence is less pronounced on the electron-doped side than on the hole-doped side. As doping is further reduced on the hole-doped but not on the electron-doped side, the momentum-space-differentiated Fermi-liquid regime gives way to a sector-selective pseudogapped regime, in which the states near $(0,\pi)$ are gapped but the states near the zone diagonal are not.  This sequence of doping dependencies, and in particular the nodal/antinodal differentiation and the momentum selective pseudogap, have for many years been understood to be key features exhibited by the high-$T_c$ copper-oxide-based superconductors. Cluster dynamical mean-field studies over the years have uncovered various aspects of this behavior. The systematic survey of different dopings and cluster sizes presented here  has firmly established the robustness of this physical picture. Nodal/antinodal momentum-space differentiation and momentum selectivity are robust features that persist independent of cluster geometry, and  there is a remarkably reliable quantitative agreement across clusters for many of the observables presented here.  We conclude that  this physics is a consequence of short-ranged correlations in the two-dimensional Hubbard model at intermediate to strong coupling.

Our comparison also allows one to understand the strengths and weaknesses of small cluster calculations. In particular, we have shown that a simple deformation of the standard $4$-site cluster designed to better separate the nodal and the antinodal region is sufficient to recover the sector-selective transition found in $8$- and $16$-site clusters. We expect that the comparative information  will be important for future use of cluster methods in more complex (multiorbital) systems, where small clusters are the only accessible ones.  

This paper has used the ``DCA'' formulation of cluster dynamical mean field theory. A different formulation, the ``Cellular Dynamical Mean Field Theory'' (CDMFT) has also been introduced \cite{Kotliar01}. CDMFT methods have not yet been applied to the range of cluster sizes and carrier concentrations needed to carry out comparisons such as those we have presented here.
Recent CDMFT papers \cite{Liebsch09,Sordi10,Sakai09} have reported results on momentum space differentiation and sector selective transitions
providing a different (but related) perspective on the issues raised in this paper.
An extension of the CDMFT results to a wider range of cluster sizes and dopings and a systematic comparison of CDMFT results to DCA results would be very desirable.

Currently, the two main limitations of cluster methods are the limited momentum-space resolution and the fermionic sign problem. These  limit the accessible cluster sizes and temperatures and  currently prevent us from reaching a definite conclusion, for example, on the precise nature of the low temperature metallic state in the nodal region in the sector-selective phase  at low temperatures or on the quantitative evolution of the Fermi arcs as a function of doping and temperature. Also while the methods give direct access to one-electron properties, wide classes of experiments involve ``two-particle'' probes whose analysis requires vertex corrections.  New theoretical developments are required to overcome these limitations.

\acknowledgments

We thank Nan Lin for providing analytical continuations of cluster data. Calculations have been performed using HPC resources from GENCI-CCRT (Grant No. 2009-t2009056112) and on the Brutus cluster at ETH Z\"urich, using a code based on the ALPS (Ref.~\onlinecite{ALPS}) libraries. A.G. and M.F. acknowledge the support of the Partner University Fund (PUF-FACE) and of the Ecole Polytechnique-EDF chair on Sustainable Energies (\'Energies Durables).  E.G. and A.J.M.  acknowledge support from the National Science Foundation Division of Materials Research under Grant No. DMR-0705847 and thank Ecole Polytechnique and the Coll\`ege de France for hospitality and travel support.  This work was initiated during a stay at the KITP where the authors were partly supported by the National Science Foundation under Grant No. NSF PHY05-51164.

\appendix

\section{Analytic Continuations \label{appendix:continuations}}

\begin{figure}[htb]
\includegraphics[width=0.9\columnwidth]{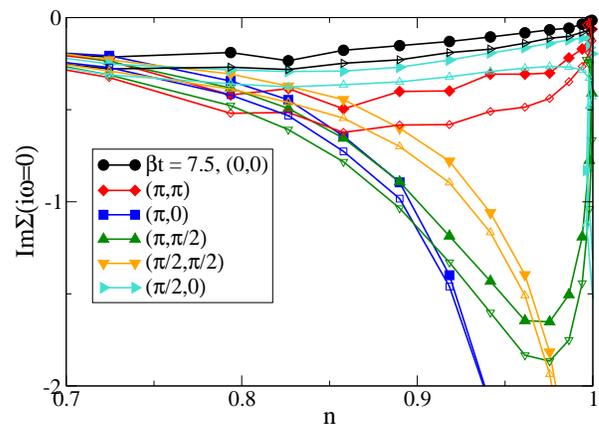}
\caption{\label{fig:sigma16site} $\Im\Sigma(\omega_n=0)$ as a function of density, for  all sectors  of $16$-site cluster, at inverse temperature $\beta=7.5/t$. Filled symbols; extrapolation based on three lowest Matsubara frequencies; open symbols; extrapolation based on two lowest Matsubara frequencies. 
}
\end{figure}

In this appendix we compare the results of the ``poor man's'' analytical continuation procedure used in the main text to maximum-entropy analytical continuation results obtained using the methods of Ref.~\onlinecite{Wang09}. We focus on the $16$-site cluster, where the restriction to high temperatures forced on us by the limits of our computational resources makes the continuations most challenging.  The solid points in  Fig.~\ref{fig:sigma16site} show $\ImSigK$  for the $16$-site cluster as a function of density, obtained by extrapolation from a fit to the three lowest Matsubara points  as described in the text.  The open points show values obtained by extrapolation from the two lowest Matsubara points.  The strong divergence associated with the $(0,\pi)$ sector is evident, as is the weaker divergence of the $(\pi/2,\pi/2)$ sector. We would like to interpret these data as indicating the presence of a growing scattering rate. However, we also see that over a wide doping range  the $(\pi,\pi/2)$ sector $\ImSigK$ is, in fact, larger than the $\ImSigK$  of the $(\pi/2,\pi/2)$ sector, even though the Fermi surface does not pass through this sector so that one might expect that low-energy processes involving this sector are suppressed. Of course, for this sector it is not correct to interpret $\ImSigK$ as a scattering rate, but the question remains of what is the meaning of the result and whether it casts doubt on the interpretation of the $\ImSigK$ in the  sectors that contain the Fermi surface.

\begin{figure}[tbh]
\includegraphics[width=0.8\columnwidth]{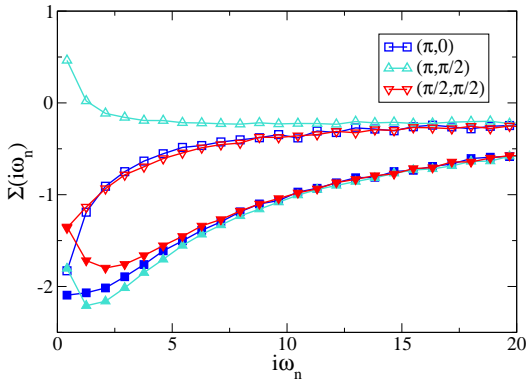}
\caption{\label{fig:sigma_mu-14} Matsubara-axis self-energy for doping $x=0.058$; solid symbols, $\text{Im} \Sigma(i\omega_n)$; open symbols $\text{Re} \Sigma(i\omega_n)$.
}
\end{figure}

To gain  further insight  we display in Fig ~\ref{fig:sigma_mu-14} the full frequency dependence of the real and imaginary parts of the  Matsubara-axis self-energy at hole doping $x=0.058$ for the sectors $(\pi/2,\pi/2)$, $(0,\pi)$, and $(\pi,\pi/2)$. We can see immediately that for all sectors  the criterion for Fermi-liquid behavior is violated. However, for the $(\pi/2,\pi/2)$ and $(0,\pi)$   sectors, both the real and imaginary parts of the self-energy behave sufficiently smoothly that a reasonable extrapolation to $\omega=0$ is possible with, in particular, answers which are reasonably consistent whether the extrapolation is performed with the two lowest Matsubara points or the three lowest. On the other hand, in the case of the $(\pi/2,\pi)$ sector both the real and imaginary parts of  the self-energy vary rapidly near $\omega=0$.

\begin{figure}[tbh]
\includegraphics[width=0.8\columnwidth]{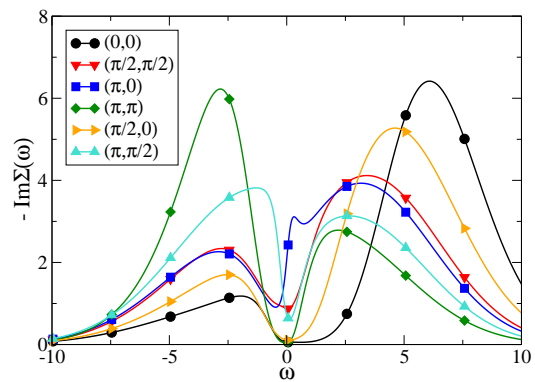}
\caption{\label{fig:imsigma16site} $\Im\Sigma(\omega)$ for $16$ site cluster at $\beta t=7.5$ and doping $x=0.058$ obtained from maximum entropy analytical continuation of self-energy data shown above. 
}
\end{figure}

The differences in Matsubara-axis self-energy are reflected in differences in the maximum entropy analytically continued self energies shown in Fig.~\ref{fig:imsigma16site}.  (Other analytically continued results are shown In Fig.~\ref{fig:sigma0beta15258site}.) We see that values of $\text{Im}\Sigma_K(\omega=0)$ for  the $K=(\pi/2,\pi/2)$ and $(0,\pi)$ sectors are approximately $-1$ and $-2$, very consistent with the values obtained from our poor-man's analytical continuation, whereas the $\text{Im}\Sigma$ in the other sectors in fact displays a gap, which is not consistent with an interpretation of the extrapolated  imaginary axis quantity as a scattering rate.  

The differences in continued behavior can be related to the structure of the Matsubara-axis data at low frequencies. In particular, we find that if extrapolations using two or three Matsubara-axis data points or three data  yield answers which are consistent with about $5\%$ and if the three lowest Matsubara points are all below the maximum in $|\text{Im} \Sigma|$ then in   the quasi-Fermi-liquid regime (as defined by $\Gamma<\pi T$) our poor man's analytical continuation  procedure yields $\ImSigk$ values which agree with analytical continuation to within about  $20-40\%$ and  $Z_{k}$'s which are about $40\%$ larger than the analytical continuation values, but with the same doping dependence. The scattering rate data we discuss in the text fulfill these conditions.  Larger differences between extrapolated and analytically continued results occur outside of the Fermi-liquid regime and for data which violate the conditions outlined above. 

\bibliography{refs_shortened}
\end{document}